\documentclass[12pt]{article}
\usepackage[margin=2.5cm]{geometry}  % add also "showframe" options  
\usepackage{epsfig, graphics}

\usepackage{subcaption}

\usepackage{caption}

\usepackage{amsfonts}
\usepackage{amssymb}
\usepackage{amsmath}
\usepackage[latin1]{inputenc}
\newcommand{\nin}{\noindent}

\def\bkR{{\rm I\kern-.17em R}}

\def \1n{1\hskip -3pt \mbox{N}}

\newfont{\bbf}{cmbx12 scaled 1435}

\begin{document}
\setlength{\baselineskip}{.26in}
\thispagestyle{empty}
\renewcommand{\thefootnote}{\fnsymbol{footnote}}
\vspace*{0cm}
\begin{center}

\setlength{\baselineskip}{.32in}
{\bbf Generalized Covariance Estimator}\\

\vspace{0.4in}

\large{ Gourieroux}\footnote{University of Toronto, Toulouse School of Economics and CREST, \\
e-mail:{\it Christian.Gourieroux@ENSAE.fr.}},
\large{ C. and J. Jasiak}\footnote{York University, e-mail: {\it jasiakj@yorku.ca}.\\
The first author gratefully acknowledges financial support of the ACPR Chair "Regulation and Systemic Risk" and the ERC DYSMOIA. The second author thanks the Natural Sciences and Engineering Council of Canada (NSERC).}

\setlength{\baselineskip}{.26in}
\vspace{0.8in}

\today\\

\medskip

\vspace{0.4in}
\begin{minipage}[t]{12cm}

\small

\begin{center}
Abstract \\
\end{center}

We consider a class of semi-parametric dynamic models with strong white noise errors. This class of processes includes the standard Vector Autoregressive (VAR) model, the nonfundamental structural VAR, the mixed causal-noncausal models, as well as nonlinear dynamic models such as the (multivariate) ARCH-M model. For estimation of processes in this class, we propose the Generalized Covariance (GCov) estimator, which is obtained by minimizing a residual-based multivariate portmanteau statistic as an alternative to the Generalized Method of Moments. We derive the asymptotic properties of the GCov estimator and of the associated residual-based portmanteau statistic. Moreover, we show that the GCov estimators are semi-parametrically efficient and the residual-based portmanteau statistics are asymptotically chi-square distributed. The finite sample performance of the GCov estimator is illustrated in a simulation study. The estimator is also applied to a dynamic model of cryptocurrency prices.

\bigskip


\end{minipage}

\end{center}
\renewcommand{\thefootnote}{\arabic{footnote}}

\newpage

\newpage

\section{Introduction}

We consider a class of semi-parametric dynamic models with strong white noise error terms. This class includes the standard Vector Autoregressive (VAR) model, the nonfundamental structural VAR, the mixed causal-noncausal models, as well as nonlinear dynamic models, such as the ARCH-M model. The assumption on the error term is used to define the Generalized Covariance (GCov) estimator of the parameter of interest by minimizing a residual-based multivariate portmanteau statistic.
The GCov estimator is introduced as a semi-parametric alternative to the GMM estimator.
This estimation approach can be applied to multivariate or univariate models. In the latter case, the objective function to be minimized can be computed from the univariate series, or from several (nonlinear) transformations of the univariate series obtained, for example, by discretizing the state space. We show that the GCov estimator is consistent and asymptotically normally distributed. 
The expression of the asymptotic variance of the GCov estimator is considerably simplified due to the underlying standardization of the criterion to be minimized. This allows us to establish  the  semi-parametric efficiency of the GCov estimator. For testing the white noise hypothesis, we  consider the residual-based multivariate portmanteau statistic. We show that this statistic follows asymptotically a chi-square distribution with the degrees of freedom adjusted for the number of identifiable parameters. This extends this well-known result, commonly used for linear dynamic VAR-type models, to a  nonlinear dynamic framework.

The following notation is used. For any $m \times n$ matrix $A$ whose $j$th column is $a_j (j= 1, . . ., n)$, $vec(A)$ will denote the column vector of dimension $mn$ defined as: 

$$vec(A) = (a_1'...,a_j',...,a_n')',$$
 
\nin where the prime denotes transposition. For any two matrices $A \equiv (a_{ij})$ and $B$, the Kronecker product $(A \otimes B)$ is the block matrix having $a_{ij}B$ for its $(i,j)$th block.

The paper is organized as follows. Section 2 recalls the interpretation  and asymptotic distribution of the multivariate portmanteau
statistic. Section 3  defines the GCov estimator and provides examples of semi-parametric models, which can be estimated by the GCov estimator. We also discuss the identifiability of the parameter of interest. Section 4 presents the results on the consistency and asymptotic normality of the GCov estimator. The residual-based multivariate portmanteau statistic and its distribution are examined in Section 5. The finite-sample performance of the GCov estimator is illustrated in Section 6 in a simulation study of ARCH-type models and mixed causal-noncausal models. The GCov estimator is also applied to the dynamic model of cryptocurrency prices. Section 7 concludes. Proofs and asymptotic expansions are gathered in Appendices 1 to 3.

\setcounter{equation}{0}\def\theequation{2.\arabic{equation}}
\section{The Weak White Noise Hypothesis}

Let us consider a univariate stationary time series $(y_t)$ with finite fourth-order moments. The test of the Weak  White Noise hypothesis $H_0= \{ \gamma(h) = 0, h=1,....,H \}$, with $\gamma(h) = Cov(y_t, y_{t-h})$ is commonly  based on the test statistic:

\begin{equation}
\xi(H) = T \sum_{h=1}^H \hat{\rho}(h)^2 = T \sum_{h=1}^H \frac{\hat{\gamma}(h)^2}{\hat{\gamma}(0)^2},
\end{equation}

\nin where $\hat{\gamma}(h)$ and $\hat{\rho}(h)$ are the sample autocovariance and autocorrelation of order $h$, respectively.

This statistic asymptotically follows a chi-square distribution $\chi^2(H)$ with $H$ degrees of freedom [see, Box, Pierce (1970)]. The aim of this Section is to review the analogue of this statistic for stationary time series of higher dimension.

Let us now consider a strictly stationary time series $(Y_t)$ of dimension $K$ with finite fourth-order moments. The null hypothesis is now $H_0 = \{ \Gamma(h) = 0, h=1,....,H \}$, where $\Gamma(h)=cov(y_t, y_{t-h})$ is the autocovariance of order $h$. The sample autocovariance is denoted by $\hat{\Gamma}(h)$ \footnote{To ensure that the sample autocovariances remain positive semi-definite: $ \hat{\Gamma}(h) = \frac{1}{T-h} (y_t -\bar{y}_t)' (y_{t-h} - \bar{y}_{t-h})$.}. The multivariate analogue of (2.1) is:

\begin{equation}
\xi(H) = T  \sum_{h=1}^H Tr [ \hat{R}^2(h)],
\end{equation} 

\nin where $\hat{R}^2(h)$ is the sample analogue of the multivariate R-square defined by:

\begin{equation}
R^2(h) =  \Gamma(h) \Gamma(0)^{-1} \Gamma(h)' \Gamma(0)^{-1}. 
\end{equation} 

\nin Since 
\begin{equation}
\hat{R}^2(h) = \hat{\Gamma}(0)^{1/2} [ \hat{\Gamma}(0)^{-1/2} \hat{\Gamma}(h) \hat{\Gamma}(0)^{-1} \hat{\Gamma}(h)'\hat{\Gamma}(0)^{-1/2}] \hat{\Gamma}(0)^{-1/2},
\end{equation}

\nin this matrix is equivalent up to a change of basis of the matrix within brackets, which is symmetric, positive-definite.  Therefore, it is diagonalisable, with a trace equal to the sum of its eigenvalues, which are the squares of the canonical correlations between $Y_t$ and $Y_{t-h}$, denoted by $\hat{\rho}_j^2(h), j=1,....,K$ [Hotelling (1936)]. Therefore:

\begin{equation}
\xi(H)  =  T \sum_{h=1}^H  Tr [ \hat{\Gamma}(h) \hat{\Gamma}(0)^{-1} \hat{\Gamma}(h)'\hat{\Gamma}(0)^{-1}] 
 =  T \sum_{h=1}^H [ \sum_{j=1}^K \hat{\rho}_j(h)^2].
\end{equation}

\nin Under the null hypothesis of Weak White Noise, this statistic follows asymptotically a chi-square distribution $\chi^2(KH)$ [see, e.g. Chitturi (1974), Robinson (1973), Anderson (1999), Section 7, Anderson (2002), Section 5].

In the special case when $\Gamma(0)$ is a diagonal matrix, formula (2.5) can be equivalently written as the sum of elements $\hat{\rho}_{ij}^2(h), i,j =1,....,K$, where $\hat{\rho}_j(h)$ are the elements of matrices of multivariate autocorrelations $\hat{D}^{-1/2} \Gamma(h) \hat{D}^{-1/2}$, where $\hat{D}$ is a diagonal matrix of variances of the component series \footnote{This approach has been used to define the GCov estimator in Gourieroux, Jasiak (2017).}.

\medskip
{\it Remark 1:} In the literature, there exist alternative test statistics that are asymptotically equivalent to the test statistic $\xi(H)$ under the null. For example, we can consider the Seemingly Unrelated Regression (SUR) model:

\begin{equation}
Y_t = \alpha + B_1 Y_{t-1}+ \cdots + B_H Y_{t-H} + u_t,
\end{equation}

\nin and introduce the statistic:

\begin{equation}
\tilde{\xi}(H) = T\; Tr [ \hat{\Gamma}^*(1) \hat{\Gamma}^*(0)^{-1} \hat{\Gamma}^*(1)'\hat{\Gamma}^*(0)^{-1} ],
\end{equation}

\nin where $\Gamma^*(1) = Cov(Y_t, \underline{Y_{t-1}}), \Gamma^*(0) = V(\underline{Y_{t-1}})$ and
$\underline{Y_{t-1}} = (Y_{t-1}',....,Y'_{t-H})'$.

Under the null, the explanatory variables in (2.6) are (asymptotically) uncorrelated, as the autoregressive coefficients $\hat{\Gamma}(h) \hat{\Gamma(0)}^{-1}$ [see, Mann, Wald (1943), Chitturi (1974), eq. (2.9)], which explains the possibility to replace the canonical correlation analysis of dimension $KH$ by $H$ canonical correlations of dimension $K$ only. We derive directly the asymptotic distribution of this statistic under the null in Appendix 1.

\setcounter{equation}{0}\def\theequation{3.\arabic{equation}}

\section{GCov Estimator}

First, we introduce a class of semi-parametric multivariate nonlinear dynamic models with independent, identically distributed (i.i.d.) errors. This i.i.d. assumption is used to construct parameter estimators by minimizing the test statistic $L_T(\theta) = \xi(H; \theta)$, corresponding to the errors and the parameter value. The maximizer of $L_T(\theta)$ is defined as the Generalized Covariance (GCov) estimator. Next, the asymptotic properties of the GCov estimator are derived.

\subsection{Semi-Parametric Model}

Let us consider a semi-parametric model satisfying dynamic equations of the type:

\begin{equation}
g(\underline{Y_t}; \theta) = u_t,
\end{equation}

\nin where $g$ is a known function, $(u_t)$ is a strong stationary white noise, and $\theta$ an unknown parameter. We assume that the model is well-specified with $\theta_0$ the true value of the parameter $\theta$. Such nonlinear structural VAR models are common in the literature, as shown in the examples below.

While the weak white noise assumption can be sufficient to define the GCov estimator and some of its asymptotic properties, the i.i.d. assumption is needed when various transformations of the series are considered and to discuss the parametric versus semi-parametric efficiency. It is also useful for the analysis of structural implementations of the models from nonlinear impulse responses [see e.g. Gourieroux, Monfort, Renne (2020), Sims(2021)]. In addition, it is commonly used in the recent literature on portmanteau tests [see e.g. Hoga(2021), Section 2].

\medskip
\nin {\bf Example 1: ARCH-M Model [Engle, Lillien, Robbins (1987)]}

The model is:

$$y_t = m(y_{t-1}; \theta_1) + \theta_3 \sigma(y_{t-1}; \theta_2) + \sigma(y_{t-1}; \theta_2) u_t.$$

\nin It extends the standard ARCH-M model by allowing for nonlinear drift and volatility functions. We have:

$$g(\underline{y_t}; \theta) = \frac{y_t - m(y_{t-1}; \theta_1) - \theta_3 \sigma(y_{t-1}; \theta_2)}{\sigma(y_{t-1}; \theta_2)}.$$

\medskip

\nin {\bf Example 2: (Causal) VAR Model}

The multivariate VAR(p) process is defined by:

$$Y_t = \Phi_1 Y_{t-1}+ \cdots + \Phi_p Y_{t-p} + u_t,$$

\nin where $\theta = [vec \Phi_1', ..., vec \Phi_p']'.$
 
We assume that the roots of the characteristic equation $det(Id - \Phi_1 \lambda - \cdots  - \Phi_p \lambda^p) = 0$
are of modulus strictly larger than one. Then, there exists a unique (strictly) stationary solution $(Y_t)$ with a causal $MA(\infty)$ representation. In this case we have:

$$g_t(Y, \theta)  =  Y_t - \Phi_1 Y_{t-1} - \cdots  \Phi_p Y_{t-p}.$$

\nin The (causal) VAR model has been widely analyzed in the literature on residual-based portmanteau test [see e.g. Hosking (1980), Li, McLeod (1981)].

\nin In the VAR specification parameters $\Phi_1,...,\Phi_p$ can be constrained. This includes in particular:

i) The VAR(1) model with reduced rank for $\Phi_1$ [see, Velu et al. (1986), Ahn, Reinsel (1988), Engle, Kozicki (1993), Reinsel, Velu (1998), Anderson (2002), Lam, Yao (2012)].

ii) The VAR(p) model with common canonical directions at all lags [see Kettenring (1971), Neuenschwander, Flury ( 1995)].

iii) The VAR(p) processes with subprocesses assumed independent [see, e.g. Haugh (1976), El Himdi, Roy (1997), Duchesne, Roy (2003), Yata, Aoshima(2016), Jin, Matteson (2018)].

iv) The VAR(p) model with causality restrictions [see e.g. Boudjellaba et al. (1994)].

v) The VAR(p) model with a Kronecker structure [Niu et al. (2020)].

\noindent and all structural VAR models. 

\medskip

\nin {\bf Example 3: Causal-Noncausal MAR(r,s) Model}
  
A multivariate $MAR(r,s)$ process [see e.g. Gourieroux, Jasiak (2017)] with mixed  causal and noncausal autoregressive dynamics 
is:

$$\Phi(L) \Psi(L^{-1}) y_t = u_t,$$

\nin where the autoregressive polynomials $\Phi(L)$ and $\Psi(L^{-1})$ are of orders $r$ and $s$, respectively.  The determinants of the autoregressive polynomials have roots outside the unit circle \footnote{The approach can be extended to the case when the polynomial cannot be factored and written as a product.}. The error $u_t$ is a non-Gaussian strong white noise process with finite first fourth moments.
Then, vector $\theta$ includes all the autoregressive matrix coefficients and function g is:

$$g_t(y, \theta)  =  \Phi(L) \Psi(L^{-1}) y_t,\; \mbox{or} \;\; g(Y_t; \theta) = \Phi(L) \Psi(L^{-1}) L^s Y_t,$$

\nin where $Y_t = y_{t+s}$. In particular, for the univariate MAR(1,1) process, with $r=s=1$:

$$ (1-\phi L)(1- \psi L^{-1}) y_t = u_t,$$

\nin where $|\phi| <1$, $|\psi| <1$, the function $g$ is easily written as:

$$g(\underline{Y_t}, \theta) = g(y_t, y_{t+1}, y_{t-1}) = Y_{t-1} - \phi \psi -\phi Y_{t-2} - \psi Y_{t},$$

\nin with $Y_t = y_{t+1}$.

\medskip The dependence of function $g$ on the past, present and future values of the observed process concerns also the application to SVAR models with nonfundamental features.

\nin {\bf Example 4: Noncausal VAR Model}

The multivariate noncausal VAR(p) process is defined by:

$$Y_t = \Phi_1 Y_{t-1}+ \cdots + \Phi_p Y_{t-p} + u_t,$$

\nin where $\theta = [vec \Phi_1', ..., vec \Phi_p']'$ and the error $u_t$ is a multivariate non-Gaussian strong white noise process with finite first fourth moments.
 
We assume that the roots of the characteristic equation $det(Id - \Phi_1 \lambda - \cdots  \Phi_p \lambda^p) = 0$
are of modulus either strictly larger, or smaller than one. Then, there exists a unique (strictly) stationary solution $(Y_t)$ with a two-sided $MA(\infty)$ representation:

$$g_t(Y, \theta)  =  Y_t - \Phi_1 Y_{t-1} - \cdots  - \Phi_p Y_{t-p}.$$

\nin The non-causal VAR(p) model has been studied in Gourieroux, Jasiak (2017) and Davis, Song (2020).

\medskip

\nin {\bf Example 5: Stacking Nonlinear Transformations of $u_t$}

From system (3.1), we can build systems of higher dimensions by considering nonlinear transformations of $u_t$. Let us introduce $J$ nonlinear transformations $a_1,...,a_J$. Then we have:

\begin{eqnarray}
a_j[g(\underline{y_t}; \theta)] & = & a_j (u_t) , \; j=1,...J, \nonumber \\
                                & \iff &  a[g(\underline{y_t}; \theta)]  = a (u_t) = v_t, 
\end{eqnarray} 

\nin where the transformed process $(v_t)$ is also a strong white noise.	

For example, the financial returns $y_t$ can be stacked together with their squares: $Y_t = (y_t, y_t^2)$ [see e.g. Wooldridge (1991), Li, Mak (1994), Ling, Li (1997), Section 4], or with their absolute values $(Y_t = y_t, |y_t|)$ [see Pena, Rodriguez (2006)], or return signs and squares $Y_t = (sign y_t, y_t^2)$ can be stacked to separate their dynamic volatility from the
bid-ask bounce effect. 

\subsection{The Estimator}

When $(u_t)$ is a strong white noise with finite moments up to order 4, $(u_t)$ is also a weak white noise. We define the GCov estimator of $\theta$ in model (3.1) by considering:

\begin{equation}
\hat{\theta}_T(H) = Arg min_{\theta} \sum_{h=1}^H Tr [ \hat{R}^2(h, \theta)],
\end{equation}

\nin where 

\begin{equation}
\hat{R}^2(h, \theta) = \hat{\Gamma}(h;\theta) \hat{\Gamma}(0, \theta)^{-1} \hat{\Gamma}(h; \theta)' \hat{\Gamma}(0;\theta)^{-1}, 
\end{equation}

\nin and $\hat{\Gamma}(h;\theta)$ is the sample covariance between $g(\underline{Y_t}; \theta)$ and $g(\underline{Y_{t-h}}; \theta)$.

Except for special cases such as the causal VAR model, where the GCov estimates of the $\Phi_j$ are equivalent to the OLS estimates, the GCov estimator has no closed-form expression.

\medskip
\nin {\it  Remark 2:} Let us consider $H=1$ and assume the parameter $\theta$ such that $dim \theta = K^2$. This is the just identified case, when $\hat{\theta}_T$ is the solution of:

$$Tr \, \hat{R}^2(1; \hat{\theta}_T) = 0 \iff \hat{R}^2(1; \hat{\theta}_T) =0 \iff \hat{\Gamma}(1; \hat{\theta}_T) = 0.$$

Thus, the GCov estimator is the analogue of a moment estimator based on centered cross-moments instead of uncentered cross-moments used in standard GMM.

\medskip
\nin {\it  Remark 3:} If $(u_t)$ has no fourth-order moment, the initial model (3.1) can be replaced by the transformed model (3.2), whenever the transformed errors $a_j(u_t)$ have such moments. Then the GCov estimator depends on $H$ and transformation $a$.

\medskip
\nin {\bf Example 6: Discretization of the State Space}

Let us consider a strictly stationary process $(y_t)$ and a discretization defined by a partition of the state space: $A_k,\; k=1,...,K+1$. We introduce the indicator functions $Y_{k,t}, \; k=1,...,K+1$, such that $Y_{kt} = 1$, if $ y_t \in A_k$, and $Y_{kt} = 0$, otherwise. Then, the transformed variables $Y_{k,t}$ have moments of any order, even if the moments of $(y_t)$ do not exist. Since $\sum_{k=1}^{K+1} Y_{k,t} = 1, \; \forall t$, we just consider
the $K$ first components to define $Y_t$. Let us denote by $p$ the K-dimensional vector with components $p_k = P[y_t \in A_k]$ and $P(h)$ the $K \times K$ matrix with elements $p_{kl}(h) = P[ y_t \in A_k, y_{t-h} \in A_l]$. We have:

$$\Gamma(h) = P(h) - pp', \; \Gamma(0) = diag p - pp' \; \mbox{and} \; \Gamma(0)^{-1} = diag (P^{-1}) - \frac{ee'}{1-e'p},$$

\nin where $e$ is the $K$-dimensional vector with unitary components. Then, it is easy to check that:

$$
\sum_{h=1}^H Tr \, R^2(h) =  \sum_{h=1}^H [ \sum_{k=1}^{K+1} \sum_{l=1}^{K+1} \frac{(p_{kl}(h) - p_k p_l)^2}{p_k p_l} ] = \sum_{h=1}^H \chi^{*2}(h),
$$

\nin where $\chi^{*2}(h)$ is the chi-square measure of (in)dependence between $Y_t$ and $Y_{t-h}$. Thus, the GCov estimator minimizes a measure of pairwise (in)dependence.

\setcounter{equation}{0}\def\theequation{4.\arabic{equation}}

\section{Asymptotic Properties of the GCov Estimator} 

Let the objective function be denoted by $L_T (\theta) = \sum_{h=1}^H Tr \, \hat{R}^2 (h; \theta)$. We provide
below the asymptotic properties of the GCov estimator. The proof of these properties are given in Appendix 2. 

\subsection{Consistency and Identification}

Under the strict stationarity of process $(Y_t)$ and the existence of the second-order moments of $g(\underline{Y_t}, \theta)$, the sample autocovariances $\hat{\Gamma}(h; \theta)$ tend to their theoretical counterparts $\Gamma(h; \theta)$ and, if $\Gamma(0; \theta)$ is invertible for $\theta \in \Theta$, then
$L_T(\theta)$ tends to:

\begin{equation}
L_{\infty}(\theta) = \sum_{h=1}^H Tr \, [R^2 (h; \theta)].
\end{equation}

\nin If model (3.1) is well-specified, we have $L_{\infty}(\theta_0) = 0$ and then the true value is the solution, which minimizes $\theta_0 = Arg min_{\theta} L_{\infty}(\theta)$. By applying the standard Jennrich's argument [Jennrich (1969), Andrews (1987)], we get the consistency of the GCov estimator under an identification condition.

\medskip

\nin {\bf Proposition 1:} If $\theta = \theta_0$ is the unique solution of the minimization of the limiting objective function $L_{\infty}(\theta)$, then the GCov estimator $\hat{\theta}_T$ is consistent.

\medskip

The uniqueness condition of the solution is an identification condition. It implies that some parameters might not be consistently approximated by a GCov estimator.

\medskip

\nin {\bf Example 6: Drift and Scale Parameters}

\nin The model (3.1) may include drift and scale parameters. In such a case:

$$g(\underline{Y_t}; \theta) = C(\tilde{g} (\underline{Y_t}; \alpha) - m),$$

\nin with $\theta=(\alpha, m, C)$. Since $Tr \,R^2$ is invariant with respect to affine transformations, $m$ and $C$ are not identifiable. Then the GCov approach has to be applied to $\tilde{g}(Y_t; \alpha)$ and parameter $\alpha$, corresponding to the identification restriction $m=0, C=Id$.

\medskip
\nin {\bf Example 7: Box-Cox transformation}

The lack of identification can also arise in nonlinear transformations. For example, let us consider the model:

\begin{equation}
Y_t (\lambda) = \frac{Y_t^{\lambda} - 1}{\lambda} = u_t,
\end{equation}

\nin where the Box-Cox transform is applied to $Y_t$, provided that it is positive. It is equivalent to assume that the $Y_t$'s are i.i.d. or that the  $Y_t(\lambda)$'s are i.i.d., then parameter $\lambda$ is not identifiable by the GCov approach.

\medskip
{\it Remark 4:} The lack of identification of $\lambda$ can be explained by considering the analogue Quasi Maximum Likelihood (QML) approach. Let us consider model (4.2) with autoregressive errors:

$$u_t = m + \Phi (u_{t-1} - m) + C \epsilon_t,$$

\nin where $(\epsilon_t)$ are i.i.d. $N(0, Id)$. The Gaussian log-likelihood has two components: the first component corresponds to the log-Jacobian of the nonlinear Box-Cox transform and the second component is the residual one. When this Gaussian log-likelihood is concentrated in $m, \Phi$ and $C$, the second component is close to $T \, Tr \, \hat{R}^2(1; \lambda)$ (when $\Phi=0$). In contrast to the GCov approach, a Gaussian QML approach identifies $\lambda$, since it accounts for the nonlinear Box-Cox transform through the log-Jacobian component. 

\medskip

If the functions $\Gamma(h; \theta), h=1,...,H$ are differentiable with respect to $\theta$ and $\theta_0$ is in the interior of the parameter set $\Theta$, the GCov estimator satisfies the first-order conditions (FOC):

\begin{equation}
\frac{\partial L_T(\hat{\theta}_T)}{\partial \theta} = 0 \iff \sum_{k=1}^K \frac{\partial \, Tr \, \hat{R}^2 ( h; \hat{\theta}_T)^2}{\partial \theta} = 0.
\end{equation}

\nin Appendix A.2.1. provides the expressions of the derivatives. We have:

\begin{eqnarray}
\frac{\partial \, Tr \, \hat{R}^2 ( h; \hat{\theta}_T)}{\partial \theta_j} & = & 2 \, Tr \, [ \hat{\Gamma}(0; \hat{\theta}_T) ^{-1} \hat{\Gamma}(h; \hat{\theta}_T)' \hat{\Gamma}(0; \hat{\theta}_T)^{-1} \frac{\partial \hat{\Gamma}(h; \hat{\theta}_T)}{\partial \theta_j}  \nonumber\\
& - & Tr\, \{ [ \hat{\tilde{R}}^2 (h; \hat{\theta}_T) \hat{\Gamma}(0; \hat{\theta}_T)^{-1} + \hat{\Gamma}(0; \hat{\theta}_T)^{-1} \hat{R}^2(h; \hat{\theta}_T)]
\frac{\partial \hat{\Gamma}(0; \hat{\theta}_T)}{\partial \theta_j} \},
\end{eqnarray}

\nin for $j=1,...,J = dim \theta$.

\medskip

\nin {\it Remark 5:} When $K=1$, the coefficient of determination is a scalar  $R^2 (h; \theta) = \rho^2(h;\theta)$ and the FOC become:

$$\sum_{h=1}^H [ \hat{\rho}(h; \theta)^2 [ \frac{ d log \hat{\gamma}(h; \theta)}{d \theta} -
\frac{ d log \hat{\gamma}(0; \theta)}{d \theta}] = 0.$$

\subsection{Asymptotic Normality and Semi-Parametric Efficiency}

Let us now expand the first-order conditions in a neighborhood of $\theta_0$. We get:

\begin{eqnarray*}
\frac{d L_T (\hat{\theta}_T)}{d \theta} & = & 0 \\
& \iff & \sqrt{T} \frac{d L_T (\theta_0)}{d \theta} + \frac{d^2 L_T (\theta_0)}{d \theta d \theta'} \sqrt{T} (\hat{\theta}_T - \theta_0) = o_p(1).
\end{eqnarray*}

\nin When $T$ tends to infinity, we have [see Appendix A.2.2]:

$$ \sqrt{T} \frac{d L_T (\theta_0)}{d \theta} = A(\theta_0) \sqrt{T} vec \hat{\Gamma}(1; \theta_0)' + o_p(1),$$
\nin and

$$ \frac{d^2 L_T (\theta_0)}{d \theta d \theta'}  = \frac{d^2 L_{\infty} (\theta_0)}{d \theta' d \theta} \equiv - J(\theta_0) + o_p(1). $$

\nin We deduce the following Proposition:

\medskip

{\bf Proposition 2:} The GCov estimator converges at speed $1/\sqrt{T}$ and we have:

$$\sqrt{T} (\hat{\theta}_T - \theta_0) = J(\theta_0) ^{-1} A(\theta_0)\sqrt{T} vec \hat{\Gamma}(1; \theta_0)' + o_p(1),$$

\nin where $J(\theta_0)$ is assumed invertible (a local  identification condition).

\medskip

\nin Since $\sqrt{T} vec \hat{\Gamma}'(1; \theta_0) $ is asymptotically normal, with mean zero, it follows that:

$$\sqrt{T} (\hat{\theta}_T - \theta_0) =N[ 0,  J(\theta_0) ^{-1} A(\theta_0) V_{asy} [\sqrt{T} vec \hat{\Gamma}(1; \theta_0)'] A(\theta_0)' J(\theta_0) ^{-1}].$$

\medskip
\medskip
\nin In general, we get a sandwich expression $J(\theta_0)^{-1} I(\theta_0) J(\theta_0)^{-1}$ of the asymptotic variance-covariance matrix of the GCov estimator [Huber (1967), White (1982)]. The expressions of matrices $J(\theta_0)$ and $A(\theta)$ are given in Appendix A.2.3., eq. (a.8), (a.9).

In our framework, the formula of the asymptotic variance-covariance matrix of the GCov estimator can be simplified [see Appendix A.2.3].

\medskip
\nin {\bf Corollary 1:} We have:

$$\sqrt{T} (\hat{\theta}_T - \theta_0) \sim N[0, \Omega(\theta_0)^{-1}],$$

\nin where

$$\Omega(\theta_0) = \sum_{h=1}^H [\frac{\partial vec \Gamma (h, \theta_0)'}{\partial \theta}
[\Gamma(0; \theta_0)^{-1} \otimes \Gamma(0; \theta_0)^{-1} ] \frac{\partial vec \Gamma (h, \theta_0)}{\partial \theta'}].$$

\nin The condition of invertibility of $\Omega(\theta_0)$ is a local identification condition:

$$Rk \, \left[ \frac{\partial vec \Gamma (1, \theta_0)'}{\partial \theta},..., \frac{\partial vec \Gamma (H, \theta_0)'}{\partial \theta} \right] = dim \, \theta, $$

\nin that ensures that the $H$ first autocovariances are locally fully informative about $\theta_0$.

\nin Corollary 1 is in particular valid for $K=1$.

\medskip

\nin {\bf Corollary 2:}  In the univariate framework $K=1$, we get:

$$\sqrt{T} (\hat{\theta}_T - \theta_0) \sim N \left\{ 0, \gamma(0; \theta_0)^2 [ \sum_{h=1}^H \frac{\partial \gamma(h; \theta_0)}{\partial \theta} \frac{\partial \gamma(h; \theta_0)}{\partial \theta'}]^{-1}\right\}.$$

The simplification in the sandwich formula (i.e. the fact that $I(\theta_0)$ is proportional to $J(\theta_0)$ with the proportionality factor given in the  proof in Appendix A.2.3) means that the GCov estimator has some degree of semi-parametric efficiency. For this reason, it is called "Generalized" by analogy to the GMM estimator [Hansen (1982)]. This efficiency is reached in a single optimization for the GCov, whereas two step optimizations are usually needed to the GMM estimator [see e.g. Lanne, Luoto (2021) for GMM estimation of structural VAR models]. This semi-parametric efficiency is a consequence of the adequate choice of weights $\hat{\Gamma}(0)^{-1}$ in the objective function [see the asymptotic behavior of $\hat{\Gamma}(h)^{-1}$ in Appendix 1].

\medskip

\nin {\it Remark 6:} The effect of $H$

The matrix $\Omega(\theta_0)$ depends on $H$. This is an increasing function of $H$ for the order on symmetric matrices: the larger $H$, the more asymptotically accurate the GCov estimator.

\medskip

\nin {\bf Example 8: Semi-Parametric versus Parametric Efficiency}

As mentioned in Example 4, the GCov approach can be applied to stacked nonlinear transformations of the series: $a(y_t)$, say. Let us consider $H=1$ for expository purpose. Then, the asymptotic accuracy of $\hat{\theta}_T$ depends on the choice of function $a$, including its dimension: $\Omega(\theta_0;a)$ depends on $a$. Then, we can expect the existence of a choice $a^*$, say, such that:

$$a^* = Arg max_a \Omega(\theta_0, a).$$ 

Let us assume that the process $(y_t)$ is a univariate Markov process and consider a grid $A_k = [a_k, a_{k+1}]$ in Example 5. When the grid step becomes very small and the number of elements in the grid increases, the chi-square measure (3.5) is equivalent to the Kullback-Leibler measure between the copula density and the uniform density. Then, the GCov estimator is a maximum likelihood estimator based on the parametric copula. Therefore, as expected $\Omega^*(\theta_0) = Max_a \Omega (\theta_0, a)$ is the information matrix corresponding to this copula-based ML approach. In other words, when $a$ varies, we reach a parametric efficiency bound for the parameters characterizing the copula. This is the well-known
property of adaptive estimation due to the i.i.d assumption on the error terms.

\setcounter{equation}{0}\def\theequation{5.\arabic{equation}}

\section{Residual-Based Multivariate Portmanteau Statistic}

The expansion of Section 4 can be used to derive the asymptotic properties of the residual-based multivariate portmanteau test statistic for $\theta$ replaced by the GCov estimator. The test statistic is:

\begin{equation}
\hat{\xi}_T (H) = L_T ( \hat{\theta}_T).
\end{equation}

\nin Let us consider the second-order expansion of $L_T(\theta_0)$, when $\theta_0$ is close to $\hat{\theta}_T$. 
As $\frac{d L_T (\hat{\theta}_T)}{ d \theta} = 0$, we get:

$$ L_T (\theta_0) \approx L_T ( \hat{\theta}_T) + \frac{1}{2} (\hat{\theta}_T - \theta_0)' \frac{d^2 L_T (\hat{\theta}_T)}{d \theta d \theta'} (\hat{\theta}_T - \theta_0),$$

\nin or,
\begin{equation}
T L_T ( \hat{\theta}_T)  \approx T  L_T (\theta_0)  - \frac{1}{2} \sqrt{T} (\hat{\theta}_T - \theta_0)' \frac{d^2 L_{\infty} (\theta_0)}{d \theta d \theta'} \sqrt{T} (\hat{\theta}_T - \theta_0).
\end{equation}

\nin The expansions of $T L_T (\theta_0)$ and $T L_T(\hat{\theta}_T)$ are derived in Appendix 3. We have:

$$
T L_T (\hat{\theta}_T) \approx  vec \,  [\sqrt{T} \hat{\Gamma} (1; \theta_0)']' \, \Pi \, vec   \, [\sqrt{T} \hat{\Gamma} (1; \theta_0)'], 
$$

\nin where the expression on $\Pi$ is given in Appendix 3, eq. (a.12). The matrix $\Pi$ is positive definite, since $T L_T(\hat{\theta}) > 0$. Then, we can apply  R.201 in Gourieroux, Monfort (1995), from which it follows that if $\Pi V_{asy} [\sqrt{T} vec \hat{\Gamma} (1; \theta_0)'] \Pi = \Pi$, then statistic $T L_T(\hat{\theta})$ follows asymptotically a $\chi^2$ distribution with the degrees of freedom equal to $K^2 - dim \theta$. We check the validity of the condition in Appendix A.3. as well as the rank of matrix $\Pi$. We get the following result:

\medskip
{\bf Proposition 3:} For any $H$ and $K$ the statistic $T L_T(\hat{\theta}_T)$ follows asymptotically the chi-square distribution: $\chi^2(K^2H - dim \theta)$.

\medskip

To understand the principle of the proof in Appendix 3, let us describe more precisely the case $K=1$ and
any $H$. The expansion of $T L_T(\hat{\theta}_T)$ becomes:

$$T L_T(\hat{\theta}_T) \sim [ \sqrt{T} \hat{\gamma}(\theta_0)]' \frac{Id-P}{\gamma(0, \theta_0)^2} [\sqrt{T} \hat{\gamma}(\theta_0)],$$

\nin with $\hat{\gamma}(\theta_0) = [ \hat{\gamma}(1;\theta_0),...,\hat{\gamma}(H;\theta_0)]'$ and $P=Z(Z'Z)^{-1}Z'$, where $Z=\partial \gamma(\theta_0)/\partial \theta'$.
As $P$ is an orthogonal projector and $V[ \sqrt{T} \hat{\gamma}(\theta_0)]= \gamma(0; \theta_0)^2 Id \equiv \Sigma$, we find that:

$$ \Pi \Sigma \Pi = \Pi.$$

\nin Therefore, we get the following corollary:

\nin {\bf Corollary 3:} For $K=1$, under the null hypothesis, the residual-based portmanteau statistic follows asymptotically a chi-square distribution with degree of freedom equal to $H$ less the rank of the matrix: 

$$\left[\frac{\partial \gamma(1; \theta_0)}{\partial \theta}, ...., \frac{\partial \gamma(H; \theta_0)}{\partial \theta}\right] = Z'.$$

\medskip

The adjustment of the degrees of freedom is equal to the rank of the Jacobian $\frac{\partial \gamma(\theta_0)}{\partial \theta'}$. When the Jacobian is of full column rank:

$$ rk \frac{\partial \gamma( \theta_0)}{\partial \theta'} = dim \theta,$$

\nin the adjustment is equal to the number of estimated parameters. This case arises when $\theta$ is identifiable by the CGov approach. This has been implicitly assumed 
when writing the inverse of matrix $J(\theta_0)$.

The result in Proposition 3 is well-established for the (causal) VAR model in Example 2, where the autoregressive parameter are estimated by the (unconstrained) OLS
[see e.g. Box, Pierce (1970), Ljung, Box (1978) for $K=1$, Chitturi (1974),  Hosking (1980), Li, McLeod (1981) for the multivariate framework].
Thus, we have extended this result to a large class of nonlinear dynamic models. This is a consequence of both an appropriate choice of the objective function and of the estimation method, and indirectly of the semi-parametric efficiency property of the GCov estimator mentioned in Section 4.

i) If the GCov estimator of $\theta$ is replaced by another estimator as a quasi-maximum likelihood estimator based on pseudo-student distribution of the $u_t$'s, say, or a nonlinear least squares estimator, the sandwich formula will not simplify and the associated residual-based portmanteau statistic will not be asymptotically chi-square distributed. For example, it may follow a mixture of chi-square distributions (see, Francq, Roy, Zakoian( 2005), theorem 3, for a special case).

ii) It has also been proposed in the literature to modify the objective criterion. For example, for the univariate setting Pena, Rodriguez (2002), (2006)  replace  $\sum_{h=1}^H Tr\, [\hat{R}^2(h)] = \sum_{h=1}^H \hat{\rho}^2(h)$, by the determinant of the Toeplitz matrix:

$$log \, det \left[ \begin{array}{ccccc} 1 & \hat{\rho}(1) & \cdots & & \hat{\rho}(H) \\
                         \hat{\rho}(1) &    \ddots          &     &   &      \vdots         \\
                         \vdots        &              &       & &               \\
                                       &              &      &  &  \hat{\rho}(1)\\
                        \hat{\rho}(H)  &              &   & \hat{\rho}(1)     & 1 
                        \end{array}   \right].
$$

An extension to a multivariate setting has been considered in Mahdi, McLeod (2012) [see also Fisher, Gallagher (2012)]. By changing the criterion, the asymptotic distribution of the residual-based portmanteau statistic can be modified. The same remark holds for a criterion such as:

$$\sum_{h=1}^H Tr \, [ \Gamma(h) \Gamma(h)'],$$

\nin [see, e.g. Lam, Yao (2012)], or $\sum_{h=1}^H Tr [[diag \, \gamma(0)]^{-1} \Gamma(h) [diag \, \gamma(0)]^{-1} \Gamma(h)']$ where $diag\, \gamma(0)$ is the diagonal matrix with terms $\gamma_{jj}(0)$ on the main diagonal [see, Forrester, Zhang (2020)].

\setcounter{equation}{0}\def\theequation{6.\arabic{equation}}

\section{Simulation Study and Application to Cryptocurrency}

The performance of GCov estimator (2) is illustrated by a simulation study. 
We consider below two semi-parametric dynamic models that are an ARCH-type model and a mixed causal-noncausal model. Next, we apply the GCov estimator to the model of cryptocurrency prices.

\subsection{ARCH-Type Model}

We consider the univariate AR(1) - ARCH(1) model:

\begin{equation}
y_t = a_0 + a_1 y_{t-1} + \epsilon_t, 
\end{equation}

\nin where $ \epsilon_ t = u_t \sigma_t $,  $\sigma_t = \sqrt{\alpha_0 + \alpha_1 \epsilon_{t-1}^2}$ and $u_t$'s are i.i.d.  Normal (0,1) distributed variables.

Parameters $a_0$ and $\alpha_0$ are fixed equal to 0 and 1, respectively, to solve the identification issue of the drift and scale effects..
We estimate parameters $a_1$ and $\alpha_1$  from samples of T=400 observations, which are replicated 200 times. 
They are estimated without constraints ensuring the stationarity conditions for process $(y_t)$.
The GCov estimator is considered with $HK^2 > dim \theta=2$, where $H=3$  and $K=2$. The two transformations applied are $\epsilon_t$ and $ |\epsilon_t|$. Tables 1a and 1b report the averaged estimated coefficients $\hat{a}$ and $\hat{\alpha}$ and their confidence intervals (CI) at level 90\%. When $a_1$ (resp. $\alpha_1$) is large, the process approaches the nonstationarity in the mean (resp. volatility persistence).

The estimators of both parameters are not significantly biased. The confidence intervals based on $\hat{a}$ do not depend on the true value of the volatility persistence. The drift parameter is easier to estimate than the volatility parameter. Thus, for some true values of the parameters, the CI for $\hat{\alpha}$ are wider than for $\hat{a}$. Finally, the estimators $\hat{a}$ and $\hat{\alpha}$ are smaller than 1, except when the volatility persistence is large (last column of Table 1b).

\newpage
{\footnotesize
\begin{center}

Table 1a AR(1)-ARCH(1) Mean Estimated $\hat{a}$ and CI at 90\%
\rotatebox{90}{
\begin{tabular}{c|cccccccccc}
\hline \\
$a$  & \multicolumn{9}{c}{$\alpha$ } \\
\hline \\
    & 0.0 & 0.1 & 0.2 & 0.3 & 0.4 & 0.5 & 0.6 & 0.7 & 0.8 & 0.9 \\
\hline \\
0.0 & 0.003 & 0.003 & -0.007 & 0.000 & -0.017 & 0.000 & -0.010 & 0.000 & -0.014 & -0.005 \\ 
CI & -0.10, 0.10 & -0.10, 0.10 & -0.10, 0.08 & -0.10, 0.08 & -0.14, 0.08 & -0.11, 0.11 & -0.14, 0.09 & -0.12, 0.10 & -0.12, 0.09 & -0.14, 0.12  \\
0.1 & 0.087 & 0.098 & 0.100 & 0.105 & 0.092 & 0.110 & 0.079 & 0.090 & 0.096 & 0.089 \\
CI & 0.00, 0.17 & 0.02, 0.18  &  0.00, 0.21 & 0.00, 0.18 & -0.01, 0.17  &  -0.00, 0.20 & -0.04, 0.18 & -0.01, 0.19 & -0.04 0.20 & 0.17, 0.21 \\
0.2 & 0.191 & 0.205 & 0.199 & 0.187 & 0.188 & 0.194 & 0.186 & 0.188 & 0.195 & 0.208 \\
CI & 0.09, 0.28 & 0.10, 0.30 & 0.09, 0.29 & 0.07, 0.28 & 0.08, 0.29 & 0.05, 0.30 & 0.04, 0.29 & 0.06, 0.31 & 0.06, 0.31 & 0.07, 0.34 \\
0.3 & 0.290 & 0.296 & 0.300 & 0.293 & 0.301 & 0.298 & 0.285 & 0.288 & 0.296 & 0.303 \\
CI & 0.20, 0.36 & 0.21, 0.37 & 0.20, 0.38 & 0.19, 0.40 & 0.20, 0.42 & 0.19, 0.42 & 0.16, 0.38 & 0.17, 0.41 & 0.17, 0.45 & 0.18, 0.40 \\
0.4 & 0.389 & 0.395 & 0.391 & 0.395 & 0.384 & 0.399 & 0.400 & 0.400 & 0.394 & 0.390 \\
CI & 0.30,  0.45 & 0.30,  0.47 & 0.29, 0.48 & 0.30, 0.48 & 0.28,  0.47 & 0.29,  0.49 & 0.29, 0.49 & 0.27, 0.51 & 0.28, 0.50 & 0.27, 0.48 \\
0.5 & 0.494 & 0.490 & 0.497 & 0.499 & 0.501 & 0.494 & 0.502 & 0.490 & 0.488 & 0.485 \\ 
CI & 0.40, 0.57 & 0.40, 0.55 & 0.40, 0.57 & 0.41, 0.57 & 0.42, 0.57 & 0.40, 0.57 & 0.39, 0.60 & 0.39, 0.57 & 0.38, 0.58 & 0.37, 0.57 \\
0.6 & 0.599 & 0.590 & 0.585 & 0.585 & 0.592 & 0.598 & 0.596 & 0.595 & 0.587 & 0.603 \\
CI & 0.50, 0.66 & 0.52, 0.65 & 0.51, 0.65 & 0.50, 0.64 & 0.49, 0.67 & 0.50, 0.71 & 0.50, 0.71 & 0.49, 0.69 & 0.50, 0.69 & 0.49 0.70 \\
0.7 & 0.671 & 0.665 & 0.669 & 0.666 & 0.670 & 0.677 & 0.676 & 0.680 & 0.691 & 0.714 \\ 
CI & 0.61, 0.73 & 0.60, 0.72 & 0.60, 0.75 & 0.59, 0.74 & 0.60,  0.75 & 0.60, 0.76 & 0.58, 0.74 & 0.58, 0.76 & 0.59, 0.88 & 0.57, 0.95 \\
0.8 & 0.810 & 0.813 & 0.810 & 0.817 & 0.818 & 0.819 & 0.829 & 0.816 & 0.818 & 0.824 \\
CI & 0.75, 0.86 & 0.75, 0.86 & 0.74, 0.87 & 0.75, 0.86 & 0.75, 0.87 & 0.75, 0.89 & 0.75, 0.94 & 0.73, 0.90 & 0.70, 0.93 & 0.72, 0.95 \\
0.9 & 0.889 & 0.888 & 0.884 & 0.888 & 0.888 & 0.890 & 0.886 & 0.896 & 0.901 & 0.908 \\ 
CI & 0.83, 0.93 & 0.82, 0.98 & 0.83, 0.95 & 0.84, 0.93 & 0.83, 0.93 & 0.83, 0.97 & 0.83, 0.95 & 0.83, 0.97 & 0.82, 0.98 & 0.84, 0.98 \\
\hline
\end{tabular}
}

\end{center}
}

\newpage
{\footnotesize
\begin{center}
Table 1b AR(1)-ARCH(1) Mean Estimated $\hat{\alpha}$ and CI at 90\%
\rotatebox{90}{
\begin{tabular}{c|cccccccccc}
\hline \\
$a$  & \multicolumn{9}{c}{$\alpha$ } \\
\hline \\
    & 0.0 & 0.1 & 0.2 & 0.3 & 0.4 & 0.5 & 0.6 & 0.7 & 0.8 & 0.9 \\
\hline \\
0.0 & 0.024 & 0.107 & 0.226 & 0.316 & 0.417  & 0.528 & 0.611 & 0.703 & 0.826 & 0.994 \\
CI & 0.00, 0.09 & 0.00, 0.25 &  0.06, 0.41 &  0.11, 0.56 & 0.19, 0.65 & 0.29, 0.84  & 0.27, 0.89 & 0.46, 0.93 & 0.48,  1.09 & 0.65, 1.52 \\
0.1 & 0.025 & 0.113 & 0.194 & 0.298 & 0.409 & 0.502 & 0.606 & 0.715 & 0.784 & 0.961 \\
CI & 0.00, 0.10 & 0.00, 0.29 & 0.05, 0.36 & 0.09, 0.51 & 0.20, 0.68 & 0.20, 0.77 & 0.30, 0.87 & 0.43, 0.97 & 0.45, 0.94 & 0.51, 1.32 \\ 
0.2 & 0.029 & 0.105 & 0.218 & 0.315 & 0.412 & 0.519 & 0.600 & 0.748 & 0.831 & 0.930 \\
CI  & 0.00, 0.12 & 0.00, 0.25  & 0.08, 0.37  & 0.09, 0.56  & 0.22, 0.63  & 0.28 0.77  & 0.32 0.89 & 0.42 0.94  & 0.49 0.96  & 0.52 1.44 \\
0.3 & 0.025 & 0.110 & 0.179 & 0.304 & 0.386 & 0.549 & 0.632 & 0.684 & 0.819  & 0.925 \\
CI & 0.00, 0.09 & 0.00, 0.25 & 0.04, 0.37 & 0.12, 0.50 & 0.19, 0.58 & 0.22, 0.78 & 0.32, 0.84 & 0.45  0.88 &  0.43, 0.90 &  0.53 1.32 \\ 
0.4 & 0.025 & 0.113 & 0.206 & 0.329 & 0.421 & 0.505 & 0.618 & 0.702 &  0.818&  0.953 \\ 
CI &  0.00, 0.12 & 0.00, 0.25 & 0.05, 0.41 & 0.14, 0.53 & 0.18, 0.67 & 0.27, 0.78 & 0.32, 0.86 & 0.41, 0.95 & 0.46, 0.94 & 0.54, 1.38 \\
0.5 &  0.034 & 0.108 &  0.218 & 0.319 & 0.418 & 0.511 &  0.605 & 0.746 & 0.813 & 0.937 \\
CI & 0.00, 0.15 & 0.00, 0.24 & 0.05, 0.39 & 0.12, 0.55 & 0.18, 0.66 & 0.28, 0.81 & 0.34, 0.89 & 0.36, 0.94 & 0.41, 0.97 & 0.54 1.50 \\ 
0.6 & 0.032 & 0.102 & 0.216 & 0.292 & 0.421 & 0.535 & 0.628 & 0.695 & 0.813 & 0.922 \\
CI & 0.00 0.12 & 0.00 0.21 & 0.06 0.45 & 0.12 0.50 & 0.21 0.70 & 0.26 0.79 & 0.37 0.88 & 0.46 0.94 & 0.49 0.96 & 0.55 1.41 \\ 
0.7 & 0.036 & 0.108 & 0.206 & 0.307 & 0.406 & 0.506 & 0.630 & 0.713 & 0.795 & 0.868 \\
CI & 0.00, 0.14 & 0.00, 0.23 & 0.05, 0.39 & 0.12, 0.48 & 0.18, 0.70 & 0.28, 0.73 & 0.32, 0.88 & 0.41, 0.94 & 0.44, 0.91 & 0.51, 1.20 \\
0.8 & 0.033 & 0.125 & 0.209 & 0.323 & 0.413 & 0.500 & 0.594 & 0.675 & 0.756 & 0.860 \\
CI & 0.00 0.10 & 0.00 0.27 & 0.07 0.37 & 0.15 0.51 & 0.19 0.64 & 0.26 0.72 & 0.36 0.82 & 0.36 0.92 & 0.41 0.94 & 0.49 0.96 \\
0.9 & 0.036 & 0.120 & 0.225 & 0.320 & 0.536 & 0.517 & 0.607 & 0.812 & 0.788 & 0.922 \\
CI & 0.00, 0.14 & 0.00, 0.24 & 0.06, 0.42 & 0.13, 0.54 & 0.17, 0.62 & 0.27, 0.74 & 0.31, 0.90 & 0.32, 0.90 & 0.47, 0.95 & 0.52, 1.42  \\
\hline
\end{tabular}
}
\end{center}
}

\medskip

\subsection{Causal-Noncausal Model}

The mixed noncausal autoregressive MAR(1,1) process is defined as the unique strictly stationary solution of:

\begin{equation}
(1- \phi L)(1-\psi L^{-1}) y_t = \varepsilon_t, 
\end{equation}

\nin where the errors are independent, identically distributed and such that $E(|\varepsilon_t|^{\delta})<\infty$ for $\delta>0$. Parameters $\phi$ and $\psi$ are two
autoregressive coefficients with modulus strictly less than one. 
Coefficient $\phi$ represents the standard causal persistence while coefficient $\psi$ depicts the noncausal persistence.

When the distribution of $\varepsilon_t$ has fat tails, the parameters acquire an additional interpretation, as a large negative error value creates a spike in the trajectory with an explosion rate of about $1/\psi$ and a collapse rate of $\phi$. We observe a jump if $\psi = 0$ and $\phi>0$ and an explosive bubble if $\psi$ is small, positive and $\phi=0$. 

When $\psi= 0$, equation (6.2) defines a purely causal autoregressive process. If $\phi= 0$, the equation above defines a purely noncausal process. If both polynomials contain non-zero coefficients, then equation (6.2) describes a mixed causal-noncausal MAR(1,1) process. The mixed process contains both leads and lags of $y_t$, and admits a two-sided moving average representation [Gourieroux, Zakoian (2015)].

As $(1-\phi L) (1-\psi L^{-1})y_t = \epsilon_t$ can also be written as $(1-(\phi L)^{-1}) (1- \psi^{-1} L)y_t = \epsilon_t/(\phi \psi)$, we introduce the constraint $\psi<1, \phi <1$ to avoid a scale identification issue.
Next, we apply the GCov estimator as an alternative to other estimation methods used in the literature on causal-noncausal models. These are the Generalized Method of Moments (GMM) method [see e.g. Lanne, Saikkonen (2011), Lanne, Luoto (2021)] and  the approximate maximum likelihood (AML) method, which is a truncated maximum likelihood [see e.g. Breidt et al. (1991), Andrews et al. (2006), Lanne, Saikkonen (2010), (2013)]. As compared with the GCov estimator, the GMM requires two steps to attain a semi-parametric efficiency, while the AML estimator assumes that the parametrized error distribution is known. Therefore, it is not robust with respect to a misspecification of the error distribution.

\medskip
We estimate the parameters $\phi$ and $\psi$ and compute the mean estimates and confidence intervals at 90\% from
100 replications of samples of size T=400 with t-student distributed errors with 6 degrees of freedom, such that $E|\epsilon|^\delta < \infty, \; \delta <6$ and the $\delta$-power moments of $y_t$ exist up to order 6. We use the simulation method of Gourieroux, Jasiak (2016).

This is a nonlinear dynamic framework with bubbles in the trajectories. Thus, it it not surprising to observe finite sample bias and rather large confidence intervals. We also observe a negative bias in $\hat{\phi}$ (resp. $\hat{\psi}$ when
$\phi < \psi$ and a positive bias, otherwise. Note that process $(y_t)$ is time reversible for $\phi=\psi$.

\newpage

\begin{center}
{\footnotesize
Table 2a MAR(1,1) Mean Estimated $\hat{\phi}$ and CI at 90\%
\rotatebox{90}{
\begin{tabular}{c|cccccccccc}
\hline \\
$\phi$  & \multicolumn{9}{c}{$\psi$ } \\
\hline \\
    & 0.0 & 0.1 & 0.2 & 0.3 & 0.4 & 0.5 & 0.6 & 0.7 & 0.8 & 0.9 \\
\hline \\
0.0 &  0.001 & 0.018 & 0.059 & 0.053 & 0.063 & 0.077 & 0.103 & 0.102 & 0.128 & 0.226 \\
CI & -0.24, 0.26 & -0.19, 0.26 & -0.21, 0.30 & -0.19, 0.34 & -0.17,  0.40 & -0.15, 0.46 &  -0.11, 0.60 &  -0.11, 0.70 &  -0.10, 0.81 & -0.09, 0.92 \\
0.1 & 0.074 & 0.104 & 0.133 & 0.138 & 0.166 & 0.170 & 0.182 & 0.199 & 0.214 & 0.199 \\
CI & -0.17,  0.30 & -0.16,  0.34 & -0.10,  0.35 & -0.11,  0.42 & -0.06,  0.45 & -0.03,  0.48 & -0.05,  0.60 & -0.03,  0.72 & -0.03,  0.82  & 0.00,  0.89 \\
0.2 &  0.141 & 0.151 & 0.172 & 0.238 & 0.257 & 0.283 & 0.298 & 0.290 & 0.305 & 0.288 \\
CI & -0.13,  0.37 & -0.07,  0.38 & -0.07,  0.42 & -0.01,  0.45  & 0.01,  0.50  & 0.04,  0.53  & 0.04,  0.64  & 0.05,  0.69  & 0.09,  0.83 &  0.09,  0.96 \\
0.3 & 0.212 & 0.242 & 0.250 & 0.294 & 0.308 & 0.356 & 0.373 & 0.378 & 0.371 & 0.349 \\
CI &  -0.11,  0.43 &  0.00,  0.47  & 0.02,  0.48  & 0.04,  0.55  & 0.09,  0.52  & 0.13,  0.63  & 0.14,  0.62  & 0.16,  0.71  & 0.18,  0.77  & 0.19,  0.87 \\
0.4 & 0.302 & 0.322 & 0.337 & 0.366 & 0.401 & 0.445 & 0.446 & 0.474 & 0.454 & 0.420  \\
CI & -0.08,  0.52  & 0.02,  0.54  & 0.08,  0.53  & 0.13,  0.58  & 0.15,  0.60  & 0.20,  0.66  & 0.20,  0.66  & 0.21,  0.76  & 0.25,  0.80  & 0.26,  0.80 \\
0.5 & 0.415 & 0.400 & 0.409 & 0.450 & 0.470 & 0.489 & 0.528 & 0.535 & 0.522 & 0.551 \\
CI & 0.04,  0.60  & 0.02,  0.63  & 0.12,  0.62  & 0.20,  0.66  & 0.22,  0.68  & 0.25,  0.71  & 0.28,  0.77  & 0.32 , 0.76  & 0.31,  0.85  & 0.36,  0.90 \\
0.6 & 0.503 & 0.488 & 0.524 & 0.511 & 0.540 & 0.573 & 0.598 & 0.627 & 0.618 & 0.659 \\
CI & 0.00,  0.70  & 0.04,  0.70  & 0.21,  0.70  & 0.20,  0.72  & 0.31,  0.74  & 0.32  0.78,  & 0.38,  0.80 &  0.37 , 0.88  & 0.41,  0.90  & 0.48,  0.96 \\
0.7 & 0.602 & 0.590 & 0.574 & 0.610 & 0.624 & 0.646 & 0.695 & 0.683 & 0.707 & 0.733 \\
CI & -0.03,  0.79 &  0.07,  0.80 &  0.12,  0.79  & 0.26,  0.81  & 0.36,  0.83 & 0.38,  0.87 & 0.45,  1.00 &  0.44,  0.97  & 0.50,  0.94  & 0.58,  0.97 \\
0.8 & 0.680 & 0.708 & 0.700 & 0.698 & 0.760 & 0.789 & 0.764 & 0.766 & 0.788 & 0.841  \\
CI & -0.04,  0.93 &  0.05,  0.91 &  0.15,  0.93 &  0.29,  1.00 &  0.29,  1.00 &  0.47,  1.00 &  0.48,  1.00 &  0.53,  1.00  & 0.59,  1.00 &  0.68,  1.00 \\
0.9 & 0.720 & 0.786 & 0.801 & 0.862 & 0.870 & 0.858 & 0.879 & 0.855 & 0.870 & 0.901 \\
CI & -0.06, 1.00 &  0.08,  1.00 &  0.16,  1.00 &  0.30,  1.00 &  0.38,  1.00 &  0.53,  1.00  & 0.60,  1.00 &  0.63 , 1.00  & 0.71,  1.00  & 0.78,  1.00 \\
 \\
\hline
\end{tabular}
}
}
\end{center}

\newpage
%\begin{center}
%Table 2b MAR(1,1) Mean Estimated $\hat{\psi}$ and CI at 90\%
{\footnotesize
\begin{center}
Table 2b: MAR(1,1) Mean Estimated $\hat{\psi}$ and CI at 90\%
\rotatebox{90}{
\begin{tabular}{c|cccccccccc}
\hline \\
$\phi$  & \multicolumn{9}{c}{$\psi$ } \\
\hline \\
    & 0.0 & 0.1 & 0.2 & 0.3 & 0.4 & 0.5 & 0.6 & 0.7 & 0.8 & 0.9 \\
\hline \\
0.0 & 0.005 & 0.075 & 0.144 & 0.242 & 0.333 & 0.418 & 0.496 & 0.598 & 0.674 & 0.686  \\
CI & -0.26,  0.25 & -0.18,  0.31 & -0.13,  0.40 & -0.04, 0.47 & -0.01,  0.53  & 0.02,  0.63 & -0.03,  0.70 & -0.04,  0.79 & -0.07,  0.88 & -0.06,  1.00 \\
0.1 & 0.018 & 0.096 & 0.159 & 0.258 & 0.325 & 0.421 & 0.505 & 0.596 & 0.691 & 0.810  \\
CI & -0.20,  0.23 & -0.13,  0.34 & -0.10,  0.38 & -0.03,  0.47  & 0.02,  0.53  & 0.08,  0.61  & 0.07,  0.69 &  0.07,  0.79  & 0.06,  1.00  & 0.09,  1.00 \\
0.2 & 0.054 & 0.144 & 0.213 & 0.259 & 0.336 & 0.410 & 0.493 & 0.603 & 0.691 & 0.825  \\
CI & -0.17,  0.30 & -0.09,  0.37 & -0.05,  0.42 &  0.03,  0.50 &  0.07,  0.55 &  0.15,  0.64  & 0.17,  0.72 &  0.19,  0.80  & 0.15,  0.89 &  0.16,  1.00 \\
0.3 & 0.081 &  0.153 & 0.240 & 0.300 & 0.381 & 0.433 & 0.519 & 0.616 & 0.724 & 0.864  \\
CI & -0.14,  0.39 & -0.08,  0.39 & -0.01,  0.47  & 0.05,  0.52  & 0.12,  0.59 &  0.17,  0.65  & 0.25,  0.72  & 0.24,  0.79  & 0.32,  0.92  & 0.34,  1.00 \\
0.4 & 0.091 & 0.174 & 0.251 & 0.327 & 0.383 & 0.444 & 0.541 & 0.615 & 0.753 & 0.879\\
CI & -0.16,  0.46 & -0.06,  0.48 &  0.02,  0.50  & 0.09,  0.57  & 0.16,  0.62  & 0.20,  0.68  & 0.30,  0.76  & 0.27, 0.83  & 0.38,  1.00  & 0.42,  1.00 \\
0.5 & 0.081 & 0.195 & 0.277 & 0.340 & 0.418 & 0.489 & 0.562 & 0.658 & 0.778 & 0.855  \\
CI & -0.14,  0.47 & -0.04,  0.56  & 0.02,  0.55  & 0.11,  0.59  & 0.20,  0.64  & 0.26,  0.70  & 0.30,  0.80  & 0.42,  0.89  & 0.44,  1.00  & 0.49,  1.00 \\
0.6 & 0.090 & 0.200 & 0.273 & 0.380 & 0.448 & 0.510 & 0.590 & 0.661 & 0.783 & 0.848  \\
CI & -0.14,  0.59 & -0.04,  0.62  & 0.05,  0.62 &  0.15,  0.68 &  0.22,  0.67 &  0.31,  0.75 &  0.36,  0.79 &  0.39,  0.91 &  0.47,  1.00 &  0.56,  1.00 \\
0.7 & 0.093 & 0.210 & 0.321 & 0.383 & 0.462 & 0.548 & 0.602 & 0.711 & 0.789 & 0.866  \\
CI & -0.12,  0.72 & -0.03,  0.73  & 0.08,  0.73  & 0.16,  0.72 &  0.21,  0.73 &  0.31,  0.77  & 0.37,  0.86  & 0.46,  1.00  & 0.54,  1.00  & 0.63,  1.00 \\
0.8 & 0.123 & 0.193 & 0.303 & 0.402 & 0.455 & 0.509 & 0.632 & 0.729 & 0.809 & 0.862  \\
CI & -0.11,  0.82 & -0.02,  0.82  & 0.06,  0.81  & 0.16,  0.80  & 0.25,  0.87  & 0.29,  0.82  & 0.40,  0.94  & 0.51,  1.00 &  0.62,  1.00 &  0.71,  1.00 \\
0.9 & 0.187 & 0.222 & 0.307 & 0.351 & 0.442 & 0.545 & 0.633 & 0.749 & 0.833 & 0.906 \\
CI & -0.11,  0.96 & -0.02,  0.96  & 0.09,  0.92  & 0.17,  0.87  & 0.28,  0.90 &  0.38,  0.86 &  0.48,  0.90 &  0.58,  1.00 &  0.68,  1.00 &  0.80,  1.00 \\
\hline
\end{tabular}
}
\end{center}
}

\subsection{Application to Bitcoin/USD Exchange Rates}

In the empirical study we consider a sample of 300 daily closing Bitcoin/USD exchange rates recorded at Yahoo Finance between 2017/07/15 and 2018/05/11. The series is plotted in Figure 1. 

\medskip

[Insert Figure 1: Bitcoin, Daily Closing Prices]
\medskip

It displays local trends, spikes and a long range of serial dependence, displayed in Figure 2, with a linear decay rate resembling a unit root process. 

\medskip

[Insert Figure 2: Bitcoin, ACF]

\medskip

The data are transformed by substracting its median of 8036.49. Next, the 
Bitcoin prices are modelled as a noncausal MAR(3,3) process with an unspecified error distribution.

\begin{equation}
(1-\phi_1 L - \phi_2 L^2 - \phi_3 L^3)(1-\psi_1 L^{-1} - \psi_2 L^{-2} - \psi_3 L^{-3})y_t = \epsilon_t,
\end{equation}

\noindent where $\epsilon_t$ is a strong white noise with a non-Gaussian distribution. The estimation results are reported below:

\begin{center}

Table 3. Estimation of MAR(3,3)

\medskip
\begin{tabular}{|c|c|c|}
\hline
parameter & estimate & st.dev  \\ \hline
  $\psi_1$ &    0.3359   &   0.0025 \\
 $\psi_2$ &  -0.0026     &  0.0033 \\
$\psi_3$ &    0.0072     &  0.0021 \\
 $\phi_1$ &     0.7029   &    0.0025  \\
$\phi_2$ &      0.1020   &    0.0031 \\
$\phi_3$ &      0.1666   &    0.0020 \\
  \hline
\end{tabular}
\end{center}

The parameter estimates are based on the series of errors and their second and third powers. The lag length $H$ is set equal to 3. The standard errors reported in Table 3 are computed from the final Hessian matrix. All estimated parameters (except for $\psi_2$) are significant. We observe some causal persistence with the sum of autoregressive coefficients equal to 0.97 and a joint bubble effect with explosive rate of about 3 $\approx \frac{1}{0.3359}$. 

The residuals are plotted in Figure 3. We observe that the residuals, unlike the Bitcoin series, do not display a trend. However, they are characterized by two regimes of low and high variance. Figure 4 shows the density of the residuals, where one can observe long tails. 

\medskip

[Insert Figure 3: MAR(3,3) Residuals]

\medskip

[Insert Figure 4: Residuals, Density]

\medskip

\noindent The model MAR(3,3) has almost removed the serial dependence in the data, as shown in the residual ACF showed in Figure 5.

\medskip

[Insert Figure 4: Residuals, ACF]

\medskip

\noindent The autocorrelation at lag 10 is quite large. However, it is difficult to comment on its significance, given the non-normal and heavy-tailed distribution of the series.

\section{Concluding Remarks}

In this paper we have introduced a semi-parametric estimation approach for a large class of nonlinear dynamic models with i.i.d. errors. The GCov estimator obtained by minimizing a multivariate portmanteau criterion has semi-parametric efficiency properties for the parameters characterizing the serial dependence. We have also shown that the associated residual-based portmanteau statistic  has asymptotically the expected chi-square distribution with an adjusted degree of freedom.

Among further extensions are the following:
1. the approach is based on the knowledge of the asymptotic behavior of the regression coefficient. This behavior is known for errors that do not have finite second order moments [see, Davis, Resnick (1986)]. In such a case the speed of convergence and asymptotic distributions of the GCov estimator are modified. The asymptotic distribution of the (residual-based) portmanteau statistic is modified as well with likely stable limiting distributions [see e.g. Gourieroux, Zakoian (2017)].

2. The set of transformations $a$  could be parametrized. For example, for financial returns we would consider the transformed process $Y_t(\lambda) = [sign(y_t), |y_t|^{\lambda}]$. Then, the GCov estimator can be defined for any $\lambda$ providing a functional estimator $\hat{\theta}(\lambda)$, indexed by $\lambda$. The distribution of this functional estimator can be derived under the null. Then, we could search for the  transformation $|y_t|^{\lambda^*}$ that is the most suitable to represent the dynamic definition of risk.

3. By considering two nested sets of  transforms of the series $y_t$, and/or different lags, we can build specification tests of the i.i.d. assumption on the errors, i.e. overidentification J-tests [see e.g. Hannan (1967), Hansen (1982), Szroeter (1983), Lanne, Luoto (2021), Section 3.3].

4. The set of transforms can also be used to focus on some features of the error distribution such as the extreme risks [see, Hoga (2021)].

\newpage
\begin{center}
REFERENCES
\end{center}

\nin Ahn, S., and G., Reinsel (1988): " Nested Reduced Rank Autoregressive Models for Multiple Time Series", JASA, 89, 849-956. 

\medskip

\nin Anderson, T. (1999): "Asymptotic Theory for Canonical Correlation Analysis", Journal  of Multivariate Analysis, 70, 1-29.
\medskip

\nin Anderson, T. (2002): "Canonical Correlation Analysis and Reduced Rank Regression is Autoregressive Models", Annals of Statistics, 30, 1134-1154.

\medskip
\nin Andrews, D. (1987): "Consistency in Nonlinear Econometric Models: A Generic Uniform Law of Large Numbers", Econometrica, 55, 1465-1471.

\medskip
\nin Andrews, B., Breidt, F, and R. Davis (2006): "Maximum Likelihood Estimation for All Pass Time Series Models", Journal of Multivariate Analysis, 97, 1638-1659.

\medskip
\nin Boudjellaba, H., Dufour, J.M. and R. Roy (1994): "Simplified Conditions for Noncausality Between Vectors in Multivariate ARMA Models", Journal of Econometrics, 63, 271-287.

\medskip
\nin Box, G. and D. Pierce (1970): "Distribution of Residual Autocorrelations in Autoregressive-Integrated Moving Average Time Series Models", JASA, 65, 1509-1526.

\medskip
\nin Breidt, J., Davis, R., Lii, K. and M. Rosenblatt (1991): "Maximum Likelihood Estimation of Noncausal Autoregressive Processes", Journal of Multivariate Analysis, 36, 175-198.

\medskip
\nin Chitturi, R. (1974): "Distribution of Residual Autocorrelations in Multiple Autoregressive Schemes", JASA, 69, 928-934.

\medskip
\nin Chitturi, R. (1976): "Distribution of Multivariate White Noise Autocorrelations", JASA, 71, 223-226.

\medskip
\nin Davis, R. and S. Resnick (1986): "Limit Theory for the Sample Covariance and Correlation Functions of Moving Averages", Annals of Statistics, 14, 533-558.

\medskip
\nin Davis, R. and L. Song (2020): "Noncausal Vector AR Processes with Application to Economic Time Series", Journal of Econometrics, 216, 246-267.

\medskip
\nin Duchesne, P. and R. Roy (2003): "Robust Tests for Independence of Two Time Series", Statistica Sinica, 13, 827-852. 

\medskip
\nin El Himdi, H. and R. Roy (1997): "Tests for Noncorrelation of Two Multivariate ARMA Time Series", Canadian Journal of Statistics, 25, 233-256.

\medskip
\nin  Engle, R. and S. Kozicki (1993): "Testing for Common Features", Journal of Business and Economic Statistics", 11, 369-395.

\medskip
\nin Engle, R. ,Lilien, D. and R. Robbins (1987): "Estimating Time Varying Risk Premia in the Term Structure: The ARCH-M Models", Econometrica, 55, 391-407.

\medskip
\nin Fisher, T. and C. Gallagher (2012): "New Weighted Portmanteau Statistic for Time Series Goodness of Fit Testing", JASA, 107, 777-787.

\medskip
\nin Forrester, P. and J. Zhang (2020): "Parametrizing Correlation Matrices", Journal of Multivariate Analysis, 178, 104619.

\medskip
\nin Francq, C., Roy, R. and J.M. Zakoian (2005): "Diagnostic Checking in ARMA Models with Uncorrelated Errors", JASA, 100, 532-544.

\medskip
\nin Gourieroux, C. and J. Jasiak (2016): "Filtering, Prediction and Simulation Methods for Noncausal Processes", Journal of Time Series Analysis, 37, 405-430. 

\medskip
\nin Gourieroux, C. and J. Jasiak (2017): "Noncausal Vector Autoregressive Process: Representation, Identification and Semi-Parametric Estimation", Journal of Econometrics, 200, 118-134.

\medskip
\nin Gourieroux, C. and A. Monfort (1995): " Statistics and Econometric Models", Vol 2, Cambridge Univ. Press.

\medskip
\nin Gourieroux, C., Monfort, A. and J.P. Renne (2020): "Identification and Estimation in NonFundamental Structural VARMA Models", Review of Economic Studies, 87, 1915-1953.

\medskip
\nin Gourieroux, C. and J.M. Zakoian (2017): "Local Explosion Modelling by Non-Causal Process", JRSS B, 79, 737-756.

\medskip
\nin Hannan, J. (1967): "Canonical Correlation and Multiple Equation Systems in Economics", Econometrica, 35, 123-138.

\medskip
\nin Hannan, J. (1976): "The Asymptotic Distribution of Serial Covariances", Annals of Statistics, 4, 396-399. 

\medskip
\nin Hansen, L. (1982): "Large Sample Properties of Generalized Method of Moment Estimators", Econometrica, 50, 1029-1054.

\medskip
\nin Haugh, L. (1976): "Checking the Independence of Two Covariance-Stationary Time Series: A Univariate Residual Cross-Correlation Approach", JASA, 71, 378-385.

\medskip
\nin Hecq, A., Lieb, L. and S. Telg (2016): "Identification of Mixed Causal-Noncausal Models in Finite Samples", Annals of Economics and Statistics, 123/124, 307-331.

\medskip
\nin Hoga, Y. (2021): "Testing for Serial Extreme Dependence in Time Series Residuals", D.P. University of Duisburg-Essen.

\medskip
\nin Horn, R. and C. Johnson (1999): "Topics in Matrix Analysis", Cambridge University Press.

\medskip
\nin Hosking, J. (1980): "The Multivariate Portmanteau Statistic", JASA, 75, 602-608.

\medskip
\nin Hosking, J. (1981)a: "Equivalent Forms of the Multivariate Portmanteau Statistic", JRSS B, 43, 261-262.

\medskip
\nin Hosking, J. (1981)b: "Lagrange Multiplier Tests of Multivariate Time Series Models", JRSS B, 43, 219-230. 

\medskip
\nin Hotelling, H. (1936) : "Relation Between Two Sets of Variants", Biometrika, 28, 321-377.

\medskip
\nin Huber, P. (1967): " The Behaviors of Maximum Likelihood Estimators Under Nonstandard Conditions", Proceeding of the Fifth Berkeley Symposium on Mathematical Statistics and Probability, Vol 1, 221-253.

\medskip
\nin Jennrich, R. (1969): "Asymptotic Properties of Nonlinear Least Squares", Annals of Mathematical Statistics, 40, 633-643.

\medskip
\nin Jin, Z. and D. Matteson (2018): "Generalizing Distance Covariance to Measure and Test Multivariate Mutual Dependence via Complete and Incomplete V-Statistics", Journal of Multivariate Analysis, 168, 304-322.

\medskip
\nin Kettenring, J. (1971): "Canonical Analysis of Several Sets of Variables", Biometrika, 58, 433-451.

\medskip
\nin Lam, C. and Q. Yao (2012): "Factor Modelling for High Dimensional Time Series", Biometrika, 98, 901-918.

\medskip
\nin Lanne, M., and J. Luoto (2021): "GMM Estimation of Non-Gaussian Structural Vector Autoregression", Journal of Business and Economic Statistics, 39, 69-81.

 \medskip
\nin Lanne, M., and P. Saikkonen (2010): "Noncausal Autoregressions for Economic Time Series", Journal of Time Series Econometrics, 3, 1-39.

\medskip
\nin Lanne, M., and P. Saikkonen (2011): "GMM Estimators with Non-Causal Instruments", Oxford Bulletin of Economics and Statistics, 71, 581-591.

\medskip
\nin Lanne, M., and P. Saikkonen (2013): " Noncausal Vector Autoregression", Econometric Theory, 29, 447-481.

\medskip
\nin Li, W. and A. McLeod (1981): "Distribution of the Residual Autocorrelations in Multivariate ARMA Time Series Models", JRSS B, 43, 231-233. 

\medskip
\nin Li, W. and T. Mak (1994): "On the Squared Residual Autocorrelations in Nonlinear Time Series with Conditional 
Heteroscedasticity", JTSA, 15, 627-639.

\medskip
\nin Lin, J. and A. McLeod (2006): "Improved Pena-Rodriguez Portmanteau Test", Computational Statistics and Data Analysis, 51, 1731-1738.

\medskip
\nin Ling, S and W. Li (1997): "On Fractionally Integrated Autoregressive Moving Average Time Series Models with Conditional Heteroscedasticity", JASA, 92, 1184-1194.

\medskip
\nin Ljung, G. and G. Box (1978): "On the Measure of Lack of Fit in Time Series Models", Biometrika, 65, 297-303.

\medskip
\nin Magnus, J. and H. Neudecker (2019): " Matrix Differential Calculus with Applications in Statistics and Econometrics", Wiley.

\medskip
\nin Mahdi, E. and I. McLeod (2012): "Improved Multivariate Portmanteau Test", Journal of Time Series Analysis", 23, 211-222.

\medskip
\nin Mann, H. and A. Wald (1943): "On the Statistical Treatment of Linear Stochastic Difference Equations", Econometrica, 11, 173-220.  

\medskip
\nin Neuenschwander, B. and B. Flury (1995): "Common Canonical Variates", Biometrika, 82, 553-560.

\medskip
\nin Niu, L., Liu, X. and J. Zhao (2020): "Robust Estimators of the Correlation Matrix with Sparse Kronecker Structure for a High Dimensional Matrix Variate", Journal of Multivariate Analysis, 177, 104598.

\medskip
\nin Pena, D. and J. Rodriguez (2002): "A Powerful Portmanteau Test of Lack of Fit for Time Series", JASA, 97, 601-610.

\medskip
\nin Pena, D. and J. Rodriguez (2006): "The Log of the Determinant of the Autocorrelation Matrix for Testing Goodness of Fit in Time Series", Journal of Statistical Planning and Inference, 136, 2706-2718.  

\medskip
\nin Reinsel, G. and R. Velu (1998): "Multivariate Reduced Rank Regressions", Springer, New York.

\medskip
\nin Robinson, P. (1973): " Generalized Canonical Analysis for Time Series", Journal of Multivariate  Analysis, 3, 141-160.

\medskip
\nin  Sims, C. (2021): "SVAR Identification through Heteroscedasticity with Misspecified Regimes", Princeton, DP.

\medskip
\nin Szroeter, J. (1983): "Generalized Wald Methods for Testing Nonlinear Implicit and Overidentifying Restrictions", Econometrica, 51, 335-353,

\medskip
\nin Velu, R., Reinsel, G. and D. Wichern (1986): "Reduced Rank Models for Multiple Time Series", Biometrika, 73, 109-118.

\medskip
\nin White, H. (1982): "Maximum Likelihood Estimation of Misspecified Models", Econometrica, 50, 1-25.

\medskip
\nin Wooldridge, J. (1991): "On the Application of Robust Regression-Based Diagnostics to Models of Conditional Means and Conditional Variances", Journal of Econometrics, 47, 5-46.

\medskip
\nin Yata, K., and M. Aoshima (2016): "High Dimensional Inference on Covariance Structures via the Extended Cross-Data Matrix Methodology", Journal of Multivariate Analysis, 151, 151-166.

\newpage

\begin{center}
\begin{figure}[h]
\centering
\includegraphics[width=12cm,angle = 0]{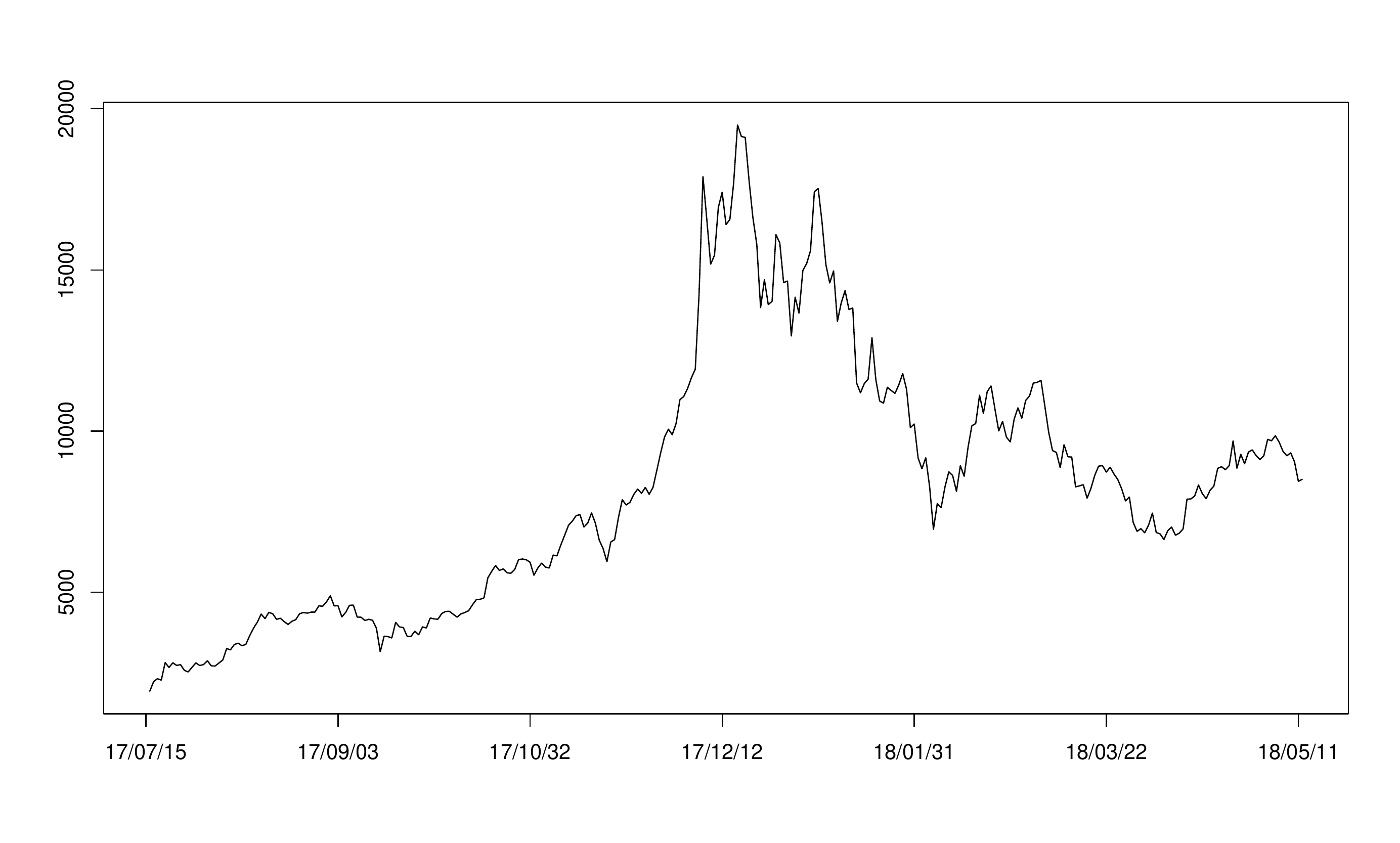}
\caption{Bitcoin, Daily Closing Prices  }
\end{figure}
\end{center}

\begin{center}
\begin{figure}[h]
\centering
\includegraphics[width=12cm,angle = 0]{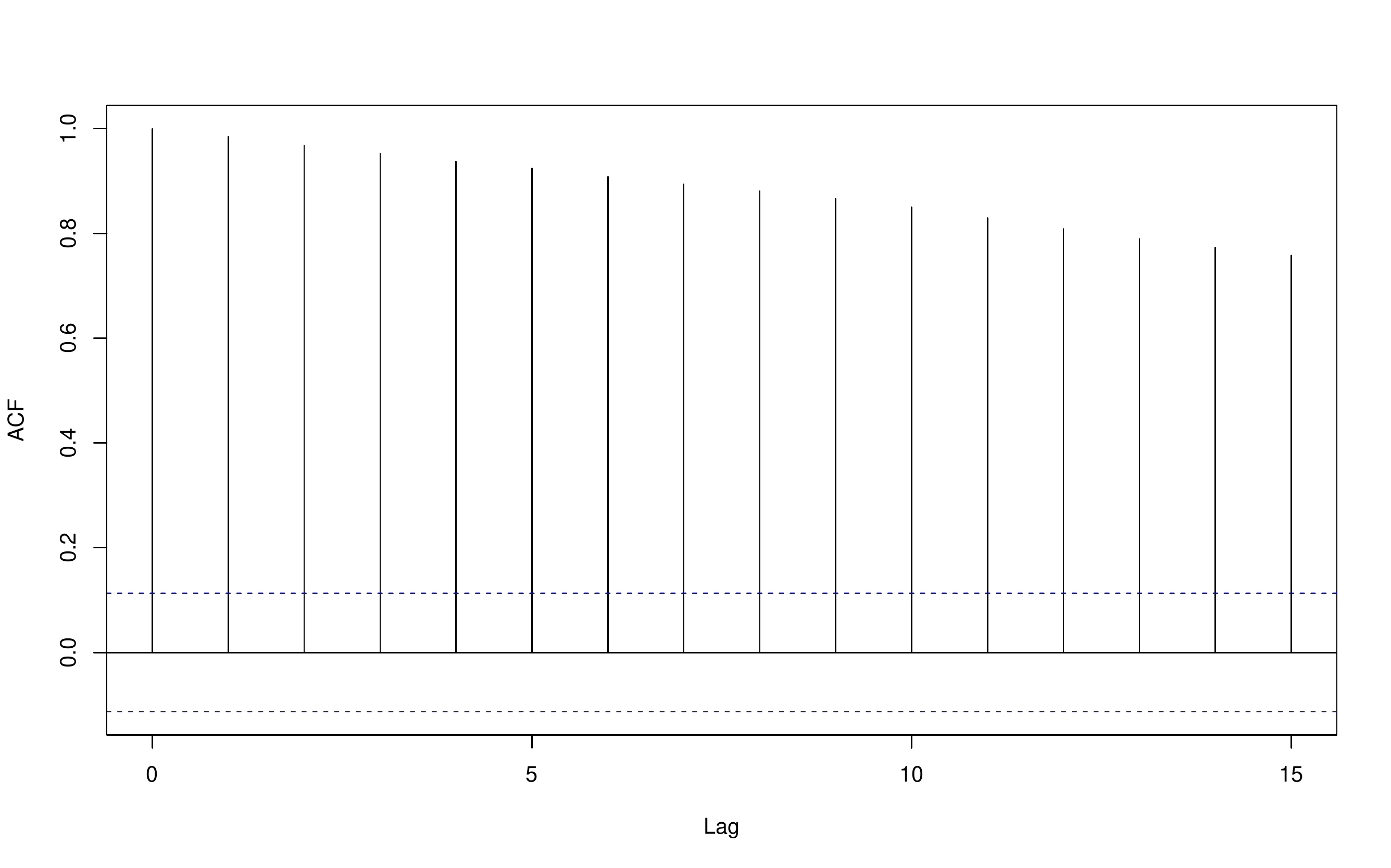}
\caption{ACF: Bitcoin}
\end{figure}
\end{center}

\newpage

\begin{center}
\begin{figure}[h]
\centering
\includegraphics[width=12cm,angle = 0]{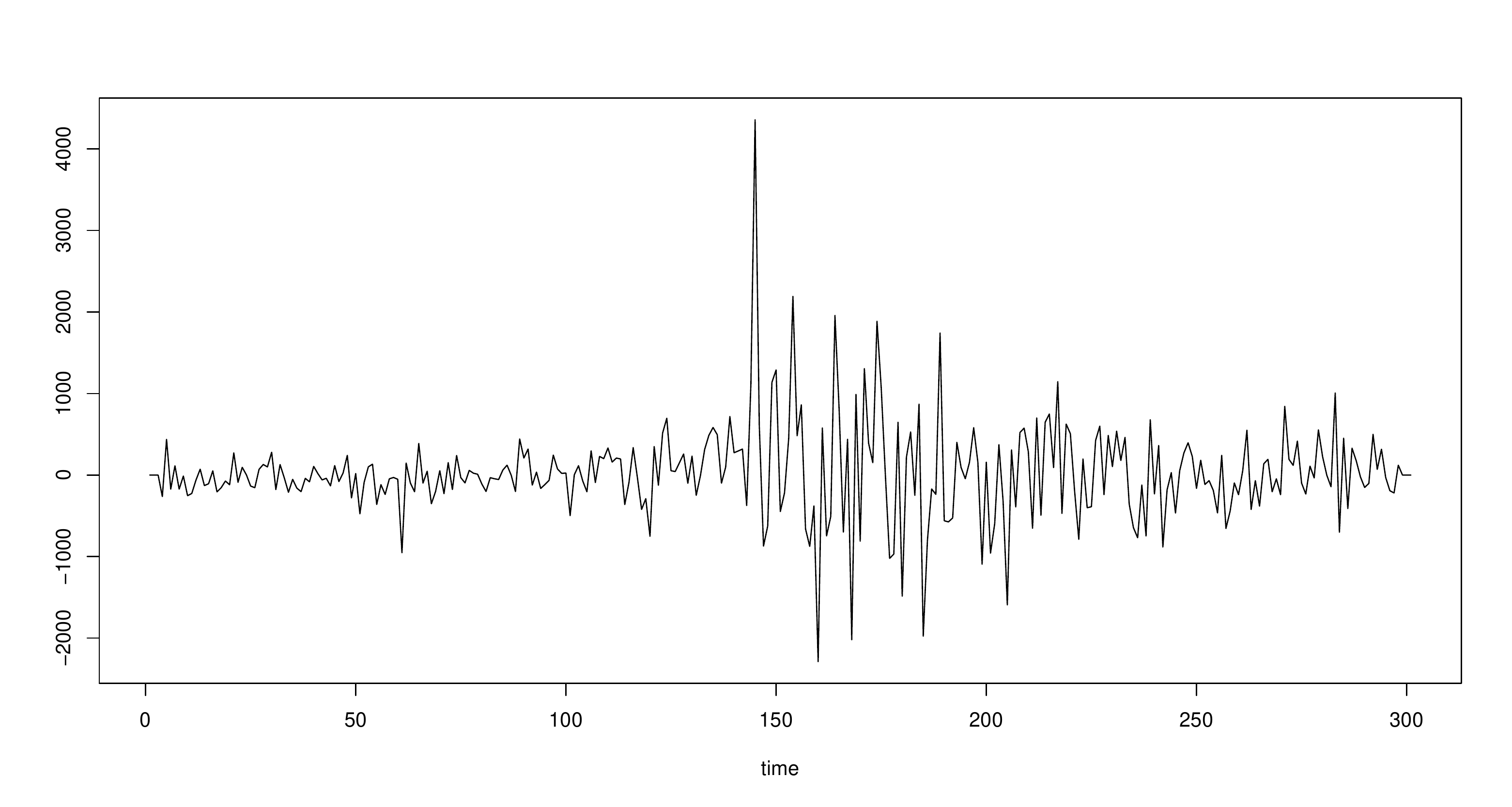}
\caption{MAR(3,3) Residuals}
\end{figure}
\end{center}

\begin{center}
\begin{figure}[h]
\centering
\includegraphics[width=12cm,angle = 0]{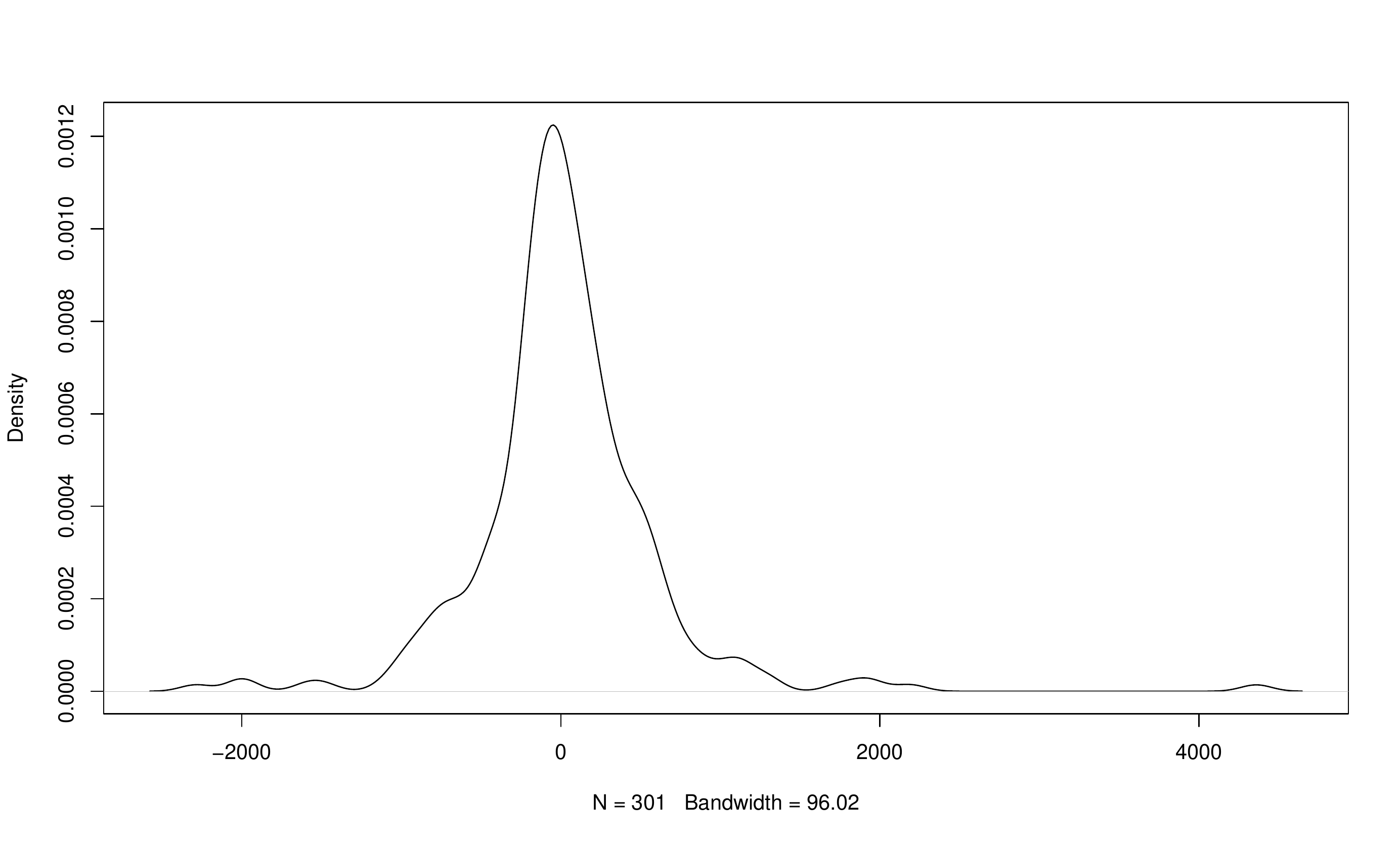}
\caption{Residuals: Density}
\end{figure}
\end{center}

\newpage

\begin{center}
\begin{figure}[h]
\centering
\includegraphics[width=12cm,angle = 0]{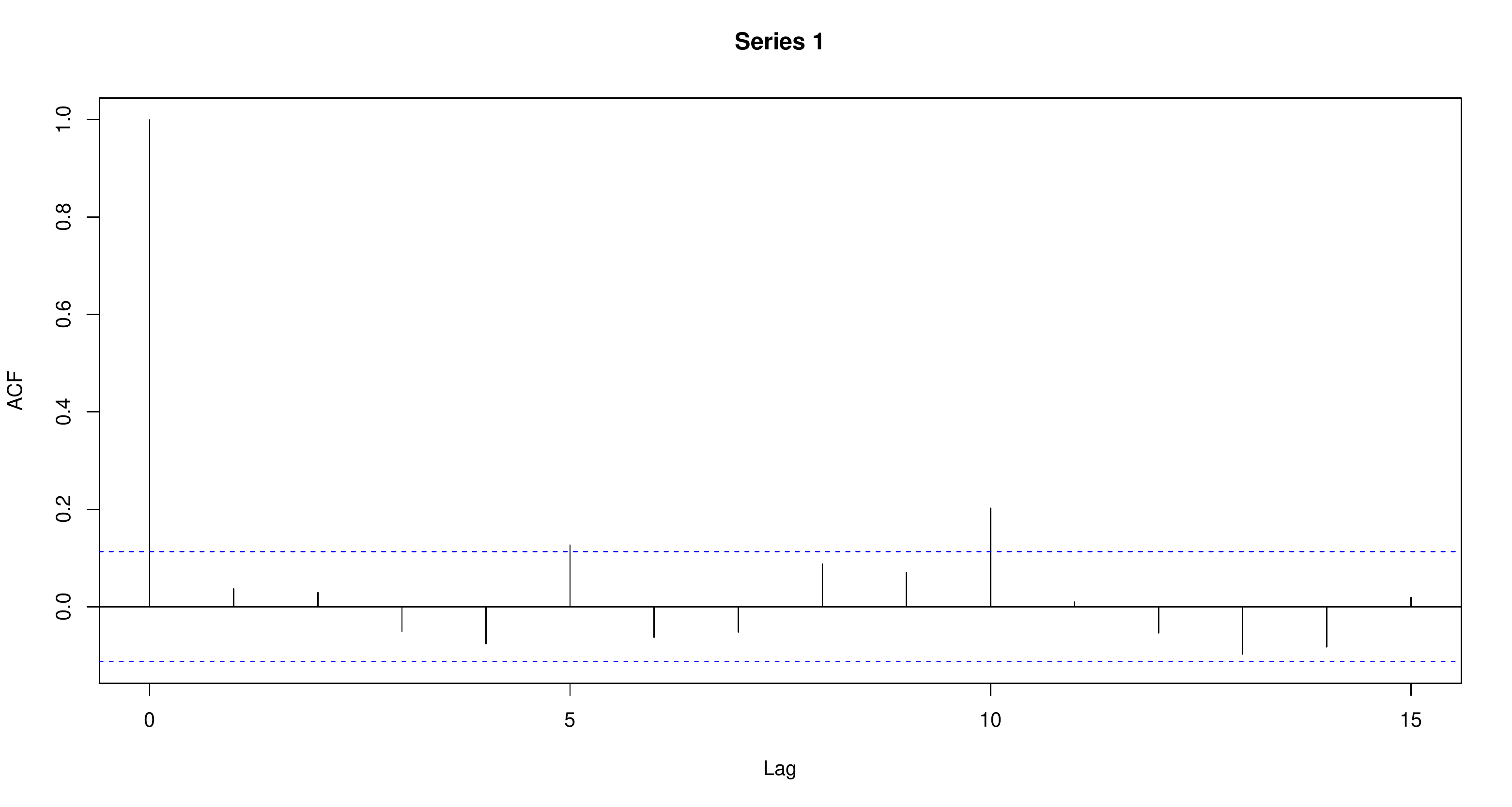}
\caption{Residuals: ACF}
\end{figure}
\end{center}

\setcounter{equation}{0}\def\theequation{a.\arabic{equation}}

\newpage
\begin{center}
APPENDIX 1 \\
{\bf The SUR Interpretation}
\end{center}

Let us consider the VAR(1) representation:

$$Y_t = \alpha + B Y_{t-1} + u_t,$$

\nin where $E(u_t)= 0$, $V(u_t) = \Sigma$, and $\Sigma$ is invertible.

This is a SUR model with identical regressors $X_t = Y_{t-1}$ in all equations. It is well-known that the GLS estimator
of $B$ is obtained by applying the OLS equation by equation. Then, we have:

$$\hat{B} = \hat{\Gamma}(1) \hat{\Gamma}(0)^{-1} \iff  \hat{B}' = \hat{\Gamma}(0)^{-1} \hat{\Gamma}(1)'.$$

\nin Moreover, we have asymptotically

\begin{equation}
\sqrt{T} [ vec (\hat{B}') - vec B'] \approx N[0, \Sigma \otimes \Gamma(0)^{-1}],
\end{equation}

\nin where the $\otimes$ denotes the Kronecker product [see e.g. Chitturi (1974),  eq. (1.13)]. In particular, under the null hypothesis $H_0 = ( \Gamma(1) = 0) = (B=0)$, we have $\Sigma= \Gamma(0)$ and

\begin{equation}
\sqrt{T} vec (B') \sim N(0, \Gamma(0) \otimes [\Gamma(0)^{-1}]).
\end{equation}

\nin Therefore, the Lagrange Multiplier test statistic \footnote{This is a Lagrange Multiplier test statistic as the asymptotic covariance matrix of $vec(\hat{B}')$ is estimated under the null hypothesis [see Hosking (1981a,b)].} for testing $H_0$ is:

\begin{eqnarray*}
\xi(1) & = & T vec [\hat{\Gamma}(0)^{-1} \hat{\Gamma}(1)']' [\hat{\Gamma}(0)^{-1} \otimes \hat{\Gamma}(0)] vec [\hat{\Gamma}(0)^{-1} \hat{\Gamma}(1)'] \\
& = & T vec [\hat{\Gamma}(0)^{-1} \hat{\Gamma}(1)']' [\hat{\Gamma}(0)^{-1/2} \otimes \hat{\Gamma}(0)^{1/2}]
[\hat{\Gamma}(0)^{-1/2} \otimes \hat{\Gamma}(0)^{1/2}]  vec [\hat{\Gamma}(0)^{-1} \hat{\Gamma}(1)']\\
& = & T vec [\hat{\Gamma}(0)^{-1/2} \hat{\Gamma}(1)'\hat{\Gamma}(0)^{-1/2} ]'vec [\hat{\Gamma}(0)^{-1/2} \hat{\Gamma}(1)'\hat{\Gamma}(0)^{-1/2} ], 
\end{eqnarray*}

\nin by using the equality: $vec(ABC) = (C'\otimes A) vec B$ [see e.g. Lemma 4.3.1 in Horn, Johnson (1999), or Magnus, Neudecker (2019), ch. 18, p. 440-441]. Moreover, we have $[vec C]'[vec C] = Tr \;CC'$. Therefore,

\begin{eqnarray}
\xi(1) & = & T \;Tr [\hat{\Gamma}(0)^{-1/2} \hat{\Gamma}(1)'\hat{\Gamma}(0)^{-1} \hat{\Gamma}(1) \hat{\Gamma}(0)^{-1/2}] \nonumber\\
& = & T \;Tr [ \hat{\Gamma}(1)' \hat{\Gamma}(0)^{-1} \hat{\Gamma}(1) \hat{\Gamma}(0)^{-1}] \nonumber\\
& = & T \;Tr \hat{R}^2(1). 
\end{eqnarray}

{\it Remark:} We easily deduce from (a.2) the asymptotic distribution of $\sqrt{T} vec[\hat{\Gamma}(1)']$ [see also Chitturi (1976), Hannan(1976)]. Indeed, we have:

$$\sqrt{T} \hat{\Gamma}(1)' = \hat{\Gamma}(0) \sqrt{T} \hat{B}' \approx \Gamma(0) \sqrt{T} \hat{B}', $$

\nin and then 
 
$$  vec[\sqrt{T} \hat{\Gamma}(1)'] = vec [ \Gamma(0) \sqrt{T} \hat{B}' ] = [Id \otimes \Gamma(0)] vec (\sqrt{T} \hat{B}' ).$$

\nin It follows that:

\begin{eqnarray*}
 vec[\sqrt{T} \hat{\Gamma}(1)'] & \approx & N[ 0, [Id \otimes \Gamma(0)] [ \Gamma(0) \otimes \Gamma(0)^{-1}] [Id \otimes \Gamma(0)]] \\
& = & N[ 0, \Gamma(0) \otimes \Gamma(0)].
\end{eqnarray*}

\newpage
\begin{center}
APPENDIX 2 \\
{\bf Asymptotic Expansions}
\end{center}

\nin {\bf  A.2.1 First-Order Derivative of $Tr \, R^2 (1; \theta)$}

\nin At lag $h=1$,  we have 
$$Tr \,R^2 (1;\theta) = Tr [\Gamma(1;\theta) \Gamma(0; \theta)^{-1} \Gamma(1; \theta)' \Gamma(0, \theta)^{-1}].$$

\nin The first-order partial derivatives are computed by considering the differential:

$$d \, Tr \, R^2(1; \theta) = \sum_{j=1}^J \frac{\partial Tr \, R^2 (1;\theta)}{\partial \theta_j} d\theta_j.$$

\nin We have:

\begin{eqnarray*}
d \, Tr \, R^2(1; \theta) & = & Tr[ d \Gamma(1;\theta) \Gamma(0; \theta)^{-1} \Gamma(1; \theta)' \Gamma(0, \theta)^{-1}] \\
& + & Tr [ \Gamma(1;\theta) d[\Gamma(0; \theta)^{-1}] \Gamma(1; \theta)' \Gamma(0, \theta)^{-1}] \\
& + & Tr [  \Gamma(1;\theta) \Gamma(0; \theta)^{-1} d\Gamma(1; \theta)' \Gamma(0, \theta)^{-1}] \\
& + & Tr[  \Gamma(1;\theta) \Gamma(0; \theta)^{-1} \Gamma(1; \theta)' d[\Gamma(0, \theta)^{-1}]],
\end{eqnarray*}

\nin because

\nin $d[ A(\theta) B(\theta)] = d A(\theta) B(\theta) + A(\theta) dB(\theta)$ and
$d \,Tr[A(\theta) +B(\theta)] = Tr \,[d A(\theta)+ d B(\theta)]$. In addition, we know that $Tr\, (AB) = Tr \,(BA), Tr\,A' = Tr \,A$. Hence,

\begin{eqnarray*}
d \, Tr \, [R^2(1; \theta)] & = & Tr \, [ \Gamma(0; \theta)^{-1} \Gamma(1; \theta)' \Gamma(0; \theta)^{-1} d \Gamma(1; \theta)] \\
& + & Tr\, [\Gamma(1; \theta)' \Gamma(0; \theta)^{-1}  \Gamma(1; \theta) d[\Gamma(0; \theta)^{-1}]]\\
& + & Tr\, [ d\Gamma(1; \theta)' \Gamma(0; \theta)^{-1} \Gamma(1; \theta) \Gamma(0; \theta)^{-1}] \\
& + & Tr \, [ \Gamma(1; \theta) \Gamma(0; \theta)^{-1} \Gamma(1; \theta)' d[\Gamma(0; \theta)^{-1}] \\
& = & 2 Tr \, [\Gamma(0; \theta)^{-1}  \Gamma(1; \theta)' \Gamma(0; \theta)^{-1} d \Gamma(1; \theta)] \\
& + & Tr \, \left[ [ \Gamma(1; \theta)' \Gamma(0; \theta)^{-1} \Gamma(1; \theta) +
\Gamma(1; \theta) \Gamma(0; \theta)^{-1} \Gamma(1; \theta)'] d[ \Gamma(0; \theta)^{-1}] \right]. 
\end{eqnarray*}

\nin We have:

$$
d[A(\theta)^{-1}] = - A(\theta)^{-1} d A(\theta) A(\theta)^{-1}.
$$

\nin Next, let us substitute this expression into the formula of $d \, Tr \, [R^2(1; \theta)]$:

\begin{eqnarray}
d \, Tr \,R^2(1; \theta) &= & 2 \,Tr \,[\Gamma(0; \theta)^{-1} \Gamma(1; \theta)' \Gamma(0; \theta)^{-1} d \Gamma(1; \theta)] \nonumber\\
& - & Tr \, \left[ \Gamma(0; \theta)^{-1} [ \Gamma(1; \theta)' \Gamma(0; \theta)^{-1} \Gamma(1; \theta) + \Gamma(1; \theta) \Gamma(0; \theta)^{-1} \Gamma(1; \theta)'] \Gamma(0; \theta)^{-1} d \Gamma(0; \theta) \right] \nonumber\\
& = & 2 Tr\, [ \Gamma(0; \theta)^{-1} \Gamma(1; \theta)' \Gamma(0; \theta)^{-1}  d \Gamma(1; \theta)] \nonumber\\
& - & Tr \, \left[ [\tilde{R}^2(1; \theta) \Gamma(0; \theta)^{-1} + \Gamma(0; \theta)^{-1}
R^2(1; \theta)] d \Gamma(0; \theta) \right], 
\end{eqnarray}

\nin with 

\begin{equation}
\tilde{R}^2(1; \theta) = \Gamma(0; \theta)^{-1} \Gamma(1; \theta)' \Gamma(0; \theta)^{-1}  \Gamma(1; \theta).
\end{equation}

\medskip
\nin {\bf A.2.2. First-Order Conditions (FOC)}

The First-Order Conditions are:

\begin{eqnarray}
\lefteqn{  \frac{ \partial Tr \, \hat{R}^2(1; \theta_j) }{ \partial \theta_j} = 0, \; j=1,...,J=dim \theta} \nonumber\\
& \iff  & 2 \,Tr [ \hat{\Gamma}(0; \theta)^{-1} \hat{\Gamma}(1; \theta)' \hat{\Gamma}(0; \theta)^{-1} 
\frac{\partial \hat{\Gamma}(1; \theta)}{\partial \theta_j}] \nonumber\\
&  & - Tr \{ \,[\hat{\tilde{R}}^2(1; \theta) \hat{\Gamma}(0; \theta)^{-1} +\hat{\Gamma}(0; \theta)^{-1} \hat{R}^2(1, \theta)] \frac{\partial \hat{\Gamma}(0; \theta)}{\partial \theta_j} \} = 0, \nonumber \\
& & \; j=1,...,J=dim \theta.
\end{eqnarray}

\medskip
\nin {\bf A.2.3. Second--Order Expansion}

We can derive the first-order conditions (FOC) in the neighborhood of $\hat{\theta} \approx \theta_0$. For some matrix function $A(\theta)$ of $\theta$, we consider the differential defined by: 

$$d A(\theta) = \sum_{j=1}^J \frac{\partial A(\theta)}{\partial \theta_j} d\theta_j, $$

\nin  with $d\theta_j  = \hat{\theta}_j - \theta_{j0}$. From (a,6), we get:
 
\begin{eqnarray}
d FOC_j (\theta) & = & 2 Tr \, \left[ \frac{\partial \hat{\Gamma}(1, \theta)}{\partial \theta_j} \,
d [\hat{\Gamma} (0; \theta)^{-1} \hat{\Gamma} (1; \theta)' \hat{\Gamma} (0; \theta)^{-1} ] \right] \nonumber\\ 
& + & 2\left[ \hat{\Gamma} (0; \theta)^{-1}   \hat{\Gamma} (1; \theta)' \hat{\Gamma} (0; \theta)^{-1} d [\frac{\partial \hat{\Gamma}(1; \theta)}{\partial \theta_j}] \right] \nonumber\\
& - & Tr \, \left[ \hat{\tilde{R}}^2 (1; \theta)  \hat{\Gamma} (0; \theta)^{-1}  + \hat{\Gamma} (0; \theta)^{-1}   \hat{R}^2(1; \theta) \right] \,   d [\frac{\partial \hat{\Gamma}(0; \theta)}{\partial \theta_j} ] \nonumber  \\
& - & Tr \, \{  [ \frac{\partial \hat{\Gamma}(0, \theta)}{\partial \theta_j}] \,
d [ \hat{\tilde{R}}^2 (1; \theta) \hat{\Gamma} (0; \theta)^{-1}  + \hat{\Gamma} (0; \theta)^{-1} 
\hat{R}^2 (1; \theta)] \}.
\end{eqnarray}

\nin In the above expression the terms without differential have to be evaluated at $\theta=\theta_0$.

\nin The expressions involving differential term $d[.]$ are linear with respect to $\hat{\theta} - \theta_0$ and then of order $1/\sqrt{T}$. Moreover, for $\theta=\theta_0$, we have $\Gamma(0; \theta_0)$ invertible, $\Gamma(1; \theta_0)=0$.
Therefore $\hat{\Gamma}(0; \theta_0)$ is of order 1 and $\hat{\Gamma}(1; \theta_0)$ of order $1/\sqrt{T}$. We deduce that the second and third components of the right hand side of the above equation (a.7) evaluated at $\theta=\theta_0$ are negligible with respect to the other components. Then, 
the right hand side of equation (a.7) can be replaced by:

\begin{eqnarray*}
\lefteqn{ 2 Tr \, \left[ \frac{\partial \hat{\Gamma}(1, \theta_0)}{\partial \theta_j} \,
d[ \hat{\Gamma} (0; \theta)^{-1} \hat{\Gamma} (1; \theta)' \hat{\Gamma} (0; \theta)^{-1}] \right] }\\
& - & Tr \left[  \frac{\partial \Gamma(0, \theta_0)}{\partial \theta_j} d
[ \hat{\Gamma} (0; \theta)^{-1} \hat{\Gamma}(1, \theta)' \hat{\Gamma} (0; \theta)^{-1} \hat{\Gamma}(1, \theta)  + \hat{\Gamma}(1, \theta) \hat{\Gamma}(0, \theta) ^{-1} 
 \hat{\Gamma}(1, \theta)' \hat{\Gamma} (0; \theta)^{-1}] \right].
\end{eqnarray*}

\medskip

\nin i) {\bf The matrix $J(\theta)$}

\nin Next, we use $d[ A(\theta) B(\theta)] = [d A(\theta)] B(\theta) + A(\theta)[ d B(\theta)]$.
Let us consider the limiting FOC, when the sample autocovariances are replaced by their theoretical counterparts and use $\Gamma(1;\theta_0)=0$. We get:

\begin{eqnarray*}
\lefteqn{
d FOC_j (\theta) = 2 Tr \, [\frac{\partial \Gamma(1; \theta_0)}{\partial \theta_j} \,
\Gamma (0; \theta_0)^{-1} d \Gamma (1; \theta)' \Gamma (0; \theta_0)^{-1} ] } \\
& - & Tr [  \frac{\partial \Gamma(0, \theta_0)}{\partial \theta_j} 
[ \Gamma (0; \theta_0)^{-1} d \Gamma(1, \theta)' \Gamma (0; \theta_0)^{-1} \Gamma(1, \theta_0) \\
& + & \Gamma(0, \theta_0)^{-1} \Gamma (1, \theta_0)' \Gamma(0, \theta_0)^{-1} d \Gamma(1, \theta) + 
  d \Gamma(1, \theta)' \Gamma (0; \theta_0)^{-1} \Gamma(1, \theta_0)' \Gamma(0, \theta_0)^{-1} \\  
 & + &  \Gamma(1, \theta_0) \Gamma (0; \theta_0)^{-1} d \Gamma(1, \theta)' \Gamma (0; \theta_0)^{-1} ]] \\
 & = & 2 \, Tr \, [\frac{\partial \Gamma(1; \theta_0)}{ \partial \theta_j} \Gamma (0; \theta_0)^{-1} d \Gamma (1; \theta)' \Gamma (0; \theta_0)^{-1}].  
\end{eqnarray*}

\nin The matrix $J(\theta_0)$ of second-order derivatives has elements $(j,k)$ such that:

$$
-J_{jk} (\theta_0)  = 2 Tr \, \left[ \frac{\partial \Gamma(1; \theta_0)}{\partial \theta_j}       
\Gamma (0; \theta_0)^{-1} \frac{\partial \Gamma(1; \theta_0)'}{\partial \theta_k} \Gamma (0; \theta_0)^{-1} \right], \; \forall j,k, 
$$

\nin or, equivalently:

\begin{equation}
J (\theta_0)  = - 2  \frac{\partial vec \Gamma(1; \theta_0)'}{\partial \theta} [\Gamma (0; \theta_0)^{-1}  \otimes   
\Gamma (0; \theta_0)^{-1} ] \frac{\partial vec \Gamma(1; \theta_0)}{\partial \theta'}. 
\end{equation}

 \medskip
\nin ii) {\bf Asymptotic equivalence for $\sqrt{T} (\hat{\theta}_T - \theta_0)$} (written for $H=1$)

\nin Hence we have:

$ \sqrt{T} [\hat{\theta}_T - \theta_0] = J(\theta_0)^{-1} \sqrt{T} \frac{d L_T (\theta)}{ d \theta} =$ $J(\theta_0)^{-1} \sqrt{T} X (\hat{\Gamma}) + o_p(1)$, where by (a.6):

\begin{eqnarray*}
\lefteqn{  X_j (\hat{\Gamma}) = 2 Tr [ \hat{\Gamma} (0; \theta_0)^{-1}  \hat{\Gamma}(1, \theta_0)' \hat{\Gamma} (0; \theta_0)^{-1} \frac{\partial \hat{\Gamma} (1, \theta_0)}{\partial \theta_j}]} \\
& - & Tr \, \{ [ \hat{\tilde{R}}^2 (1; \theta_0) \hat{\Gamma} (0; \theta_0)^{-1} + \hat{\Gamma} (0; \theta_0)^{-1}  \hat{R}^2 (1; \theta_0)] \frac{\partial \hat{\Gamma} (1, \theta_0)}{\partial \theta_j}\},
\end{eqnarray*}

\nin by (a.6). It is Normally distributed as $\hat{\Gamma} \approx \Gamma$ with the Central Limit Theorem that holds for the sample autocovariances. 

We have $X_j (\hat{\Gamma}) - X_j (\Gamma) = d  X_j (\Gamma)$. We can disregard all terms including $\Gamma(1, \theta_0)$ as $\Gamma(1; \theta_0) = 0$. Hence,

\begin{eqnarray*}
\sqrt{T} X_j (\hat{\Gamma}) & \approx & 2 Tr \, [ \Gamma(0; \theta_0)^{-1} \sqrt{T} \hat{\Gamma} (1; \theta_0)
\Gamma(0; \theta_0)^{-1} \frac{\partial \Gamma'(1; \theta_0)}{\partial \theta_j}]\\
& \approx & 2 Tr \, [ \Gamma(0; \theta_0)^{-1} \frac{\partial \Gamma'(1; \theta_0)}{\partial \theta_j} \Gamma(0; \theta_0)^{-1} \sqrt{T} \hat{\Gamma}(1; \theta_0)],
\end{eqnarray*}

\nin by applying $Tr \, A' = Tr \, A$ and $Tr (AB)= Tr(BA)$.

Since $Tr\, (A'B) = ( vec A)'(vec B)$ and $vec (A \otimes C) = (C' \otimes A)$, we deduce:

\begin{eqnarray*}
\sqrt{T} X_j (\hat{\Gamma}) & \approx & 2 vec [\Gamma(0; \theta_0)^{-1} \frac{\partial \Gamma(1; \theta_0)}{\partial \theta_j} \Gamma(0; \theta_0)^{-1}]' vec [\sqrt{T} \hat{\Gamma}'(1; \theta_0)] \\
& = & A_j(\theta) vec [ \sqrt{T} \hat{\Gamma}'(1; \theta_0)], \; j=1,...,J=dim \theta.
\end{eqnarray*}

\nin or,

$$\sqrt{T} X(\hat{\Gamma}) = A(\theta) vec[ \sqrt{T} \hat{\Gamma}'(1; \theta_0)],$$

\nin where

\begin{equation}
A(\theta) = 2 \frac{\partial vec \Gamma(1; \theta)'}{\partial \theta} \left[\Gamma(0; \theta_0)^{-1} \otimes \Gamma(0; \theta_0)^{-1}\right].
\end{equation}

\nin From the remark in Appendix 1, it follows that:

$$vec [\sqrt{T} \hat{\Gamma}'(1;\theta_0)] \approx N[0, \Gamma(0;\theta_0) \otimes \Gamma(0;\theta_0)].$$

\nin Therefore $\sqrt{T} X (\hat{\Gamma})$ is asymptotically normally distributed, with mean zero and asymptotic variance:

\begin{equation}
V_{asy}[\sqrt{T} X (\hat{\Gamma})] = - 2 J(\theta_0).
\end{equation}

\nin From (a.10), and the absence of correlation between the matrices $\hat{\Gamma}(h)$ at different lags, we deduce the simplification in deriving the asymptotic variance-covariance matrix of $\sqrt{T} (\hat{\theta}_T - \theta_0)$ in Corollary 1.

\newpage
\begin{center}
APPENDIX 3 \\
{\bf Expansion of the Multivariate Portmanteau Statistic}
\end{center}

\nin (i) Let us consider the asymptotic expansion of: 
$$
T L_T(\theta_0) = T \,Tr\, [ \hat{\Gamma}(1; \theta_0) \hat{\Gamma}(0; \theta_0)^{-1} \hat{\Gamma}(1; \theta_0)' \hat{\Gamma}(0; \theta_0)^{-1}], $$

\nin written for $H=1$. Since $ Tr (C'C) = [vec\, C]'[vec\, C]$, we have:
$$
T L_T (\theta_0) = T vec [  \hat{\Gamma}(1; \theta_0)' \hat{\Gamma}(0; \theta_0)^{-1} ]' vec [\hat{\Gamma}(1; \theta_0)' \hat{\Gamma}(0; \theta_0)^{-1} ].$$

\nin Moreover, since $vec (ABC) = (C' \otimes A) vec B$, we get:

\begin{eqnarray*}
T L_T (\theta_0) & \approx & [ \sqrt{T} vec \hat{\Gamma}'(1; \theta_0)]' [\Gamma(0;\theta_0)^{-1} \otimes \Gamma(0;\theta_0)^{-1}]
 [ \sqrt{T} vec \hat{\Gamma}'(1; \theta_0)].
\end{eqnarray*}

\nin (ii) Then, let us consider the asymptotic expansion (5.2) of the standardized residual-based portmanteau statistic. By using (a.8), (a.9) and the above result, we get:

\begin{equation}
T L_T (\hat{\theta}_T) \approx vec [ \sqrt{T} \hat{\Gamma}'(1; \theta_0)']'\, \Pi\, vec [ \sqrt{T} \hat{\Gamma}'(1; \theta_0)'],
\end{equation} 

\nin with

\begin{eqnarray}
\Pi &  = & \Gamma(0;\theta_0)^{-1} \otimes \Gamma(0;\theta_0)^{-1} \nonumber \\
& - & [ \Gamma(0;\theta_0)^{-1} \otimes \Gamma(0;\theta_0)^{-1}] \frac{\partial vec \Gamma(1; \theta_0)}{\partial \theta'} \nonumber \\
& & \left\{ \frac{\partial vec \Gamma(1; \theta_0)'}{\partial \theta}  [ \Gamma(0;\theta_0)^{-1} \otimes \Gamma(0;\theta_0)^{-1} ]    \frac{\partial vec \Gamma(1; \theta_0)}{\partial \theta}       \right\}^{-1} \nonumber \\
& & \frac{\partial vec \Gamma(1; \theta_0)'}{\partial \theta} [ \Gamma(0;\theta_0)^{-1} \otimes \Gamma(0;\theta_0)^{-1} ].
\end{eqnarray}

\nin (iii) The condition

$$ \Pi \, V_{asy} [\sqrt{T} \, vec \hat{\Gamma}(1; \theta_0)'] \, \Pi = \Pi,$$

\nin with $V_{asy} \sqrt{T}\, vec [\hat{\Gamma}(1; \theta_0)'] = \Gamma(0; \theta_0) \otimes  \Gamma(0; \theta_0)$
is verified. It is due to an interpretation in terms of orthogonal projection of $vec[ \sqrt{T} \hat{\Gamma}'(1; \theta_0)']$ on the $\partial vec \Gamma(1; \theta_0)'/\partial \theta$ for the scalar product associated with $\Gamma(0;\theta_0) \otimes \Gamma(0;\theta_0)$.

\end{document}